\documentclass[11pt]{article}
\usepackage{jheppub}
\usepackage{bm,bbm}
\usepackage{booktabs,amsmath}
\usepackage{accents}
\usepackage{multirow}
\usepackage{graphics}
\usepackage{braket}
\usepackage{mathtools}
\usepackage{comment}
\usepackage{autobreak}
\usepackage{subcaption}
\usepackage{enumerate}
\usepackage{longtable}
\usepackage{enumitem}

\usepackage{nicematrix}

\usepackage{tikz}

\usetikzlibrary{patterns}
\usetikzlibrary{arrows,shapes,positioning}
\usetikzlibrary{decorations.markings}
\usetikzlibrary{decorations.pathmorphing}
\tikzset{snake it/.style={decorate, decoration=snake}}

\hypersetup{
    colorlinks=true,
    linkcolor=blue,
    filecolor=magenta,      
    urlcolor=cyan,
    pdfpagemode=FullScreen,
}

\usepackage{arydshln}
\usepackage{mathtools}
\edef\restoreparindent{\parindent=\the\parindent\relax}
\usepackage{parskip}
\restoreparindent
\usepackage{ascmac}
\usepackage{mathrsfs}

\renewcommand{\arraystretch}{1.25}

\usepackage{amsthm}
\newtheoremstyle{break}
  {\topsep}{\topsep}%
  {\upshape}{}%
  {\bfseries}{}%
  {\newline}{}%
\theoremstyle{break}

\def\Tr{{\rm Tr}}
\def\d{{\rm d}}
\def\i{{\rm i}}
\def\CB{{\cal B}}
\def\CC{{\cal C}}

\def\CF{{\cal F}}
\def\CG{{\cal G}}
\def\CH{{\cal H}}

\def\CL{{\cal L}}

\def\CN{{\cal N}}
\def\CO{{\cal O}}
\def\CP{{\cal P}}

\def\CS{{\cal S}}

\def\CZ{{\cal Z}}
\def\BC{\mathbb{C}}
\def\BF{\mathbb{F}}

\def\BR{\mathbb{R}}

\def\BZ{\mathbb{Z}}

\def\Ba{\mathbf{a}}
\def\Bb{\mathbf{b}}

\def\SH{\mathsf{H}}
\def\SW{\mathsf{W}}
\def\d{\mathrm{d}}
\def\SU{\mathrm{SU}}

\def\U{\mathrm{U}}

\title{
Quantum subsystem codes, CFTs and their $\BZ_2$-gaugings}

\author[a]{Keiichi Ando,} 
\author[b]{Kohki Kawabata,}
\author[a]{and Tatsuma Nishioka}

\affiliation[a]{Department of Physics, Osaka University,\\
Machikaneyama-Cho 1-1, Toyonaka 560-0043, Japan}

\affiliation[b]{Department of Physics, Faculty of Science,
The University of Tokyo,\\
Bunkyo-Ku, Tokyo 113-0033, Japan}

\preprint{OU-HET-1233}

\abstract{
We construct Narain conformal field theories (CFTs) from quantum subsystem codes, a more comprehensive class of quantum error-correcting codes than quantum stabilizer codes, for qudit systems of prime dimensions.
The resulting code CFTs exhibit a global $\BZ_2$ symmetry, enabling us to perform the $\BZ_2$-gauging to derive their orbifolded and fermionized theories when the symmetry is non-anomalous.
We classify a subset of these subsystem code CFTs using weighted oriented graphs and enumerate those with small central charges.
Consequently, we identify several bosonic code CFTs self-dual under the $\BZ_2$-orbifold, new supersymmetric code CFTs, and a few fermionic code CFTs with spontaneously broken supersymmetry.
}

\begin{document} 
\maketitle
\flushbottom

\newpage

\section{Introduction}
Quantum error-correcting theory (QEC) has emerged as a pivotal framework to preserve and manipulate quantum information with high fidelity \cite{Shor:1995hbe,Steane:1996va,Knill:1996ny,Gottesman:1997zz,nielsen2002quantum}. 
Initially developed to address the inherent fragility of quantum states in quantum computing, QEC has transcended its original domain in quantum information science, finding profound applications in condensed matter theory \cite{Kitaev:1997wr,Levin:2004mi,Bullock:2006bv,Chamon:2004lew,Haah:2011drr,Vijay:2015mka,Vijay:2016phm} and high energy physics \cite{Almheiri:2014lwa,Pastawski:2015qua,Pastawski:2016qrs,Hayden:2016cfa}.
This interdisciplinary proliferation highlights the versatility of QEC principles and their fundamental relevance to modern physics.

Recently, a novel construction of two-dimensional conformal field theories (CFTs), called code CFTs, from quantum stabilizer codes has been discovered \cite{Dymarsky:2020qom,Kawabata:2022jxt,Dymarsky:2021xfc,Alam:2023qac}, which has triggered further investigations into CFTs of Narain type \cite{Narain:1985jj,Narain:1986am} from the viewpoint of (quantum) error-correcting theory \cite{Yahagi:2022idq,Angelinos:2022umf,Furuta:2022ykh,Kawabata:2023usr,Kawabata:2023iss,Furuta:2023xwl}.
A sample of the diverse applications of code CFTs can be found in \cite{Dymarsky:2020pzc,Dymarsky:2020bps,Buican:2021uyp,Henriksson:2022dnu,Henriksson:2022dml,Dymarsky:2022kwb,Buican:2023bzl,Aharony:2023zit,Barbar:2023ncl,Buican:2023ehi}.
See also \cite{frenkel1989vertex,Dixon:1988qd,Dolan:1994st,Gaiotto:2018ypj,Henriksson:2021qkt,Kawabata:2023nlt,Kawabata:2023rlt,Moore:2023zmv} for similar constructions of chiral CFTs from classical codes.

In this paper, we construct code CFTs from quantum subsystem codes, the most versatile class of QEC codes with greater fault tolerance and resilience against noise compared to quantum stabilizer codes \cite{poulin2005stabilizer,kribs2005operator,Bacon:2006nse,aliferis2007subsystem,aly2006subsystem,aly2008subsystem,aly2009constructions}.
A subsystem code is a Hilbert space of an $n$ qudit system with the following structure:
\begin{align}\label{subsystem_hilbert_space}
    \CH = (\CH_L \otimes \CH_G)\oplus \CH_R \ ,
\end{align}
where $\CH_L$ is the Hilbert space encoding quantum information of $k$ ``logical" 
 qudits, $\CH_G$ the ``gauge" subsystem of $r$ qudits which is not necessarily protected against noise, and $\CH_R$ the rest of $\CH$, respectively.
The structure \eqref{subsystem_hilbert_space} of the Hilbert space is characterized by the gauge group $\CG$, a non-abelian subgroup of the $n$-qudit Pauli group.
Subsystem codes of $n$ qudits with $k$ logical and $r$ gauge qudits are denoted by $[[n,k,r]]$.
Quantum stabilizer codes are subsystem codes without gauge qudits, whose gauge groups are abelian, and denoted by $[[n,k]]$ if they have $k$ logical qudits. 

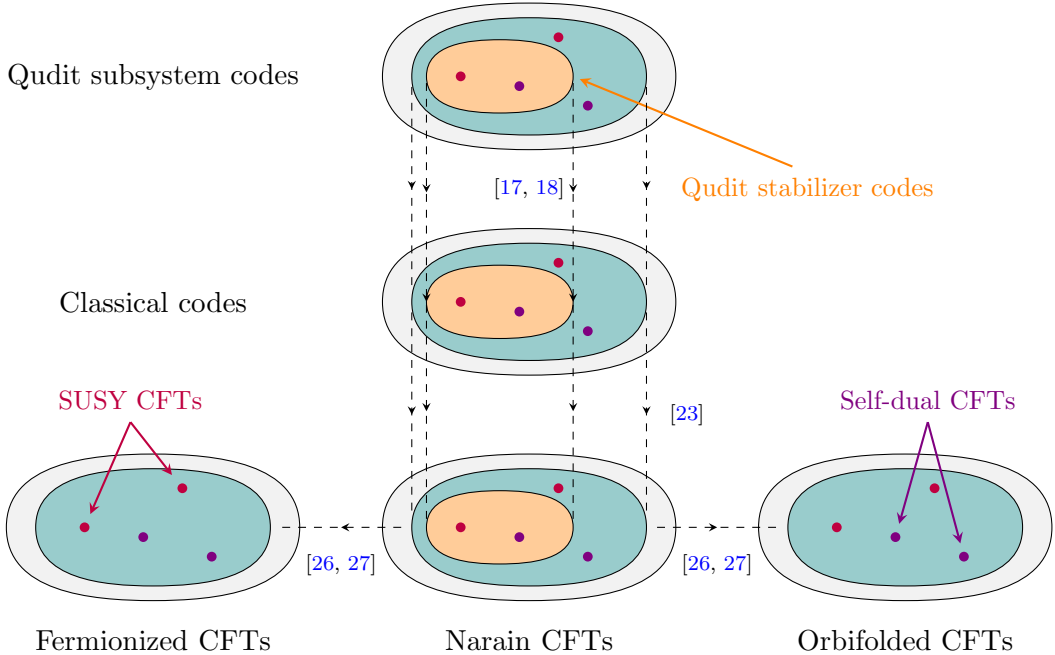
\begin{figure}[hp]
    \begin{center}
        \begin{tikzpicture}[transform shape, >=stealth]
        \begin{scope}[yshift=2.5cm]
            \begin{scope}[scale=1.3]
                \filldraw[fill=lightgray!20] (0,-0.75) to[out=180,in=-90] (-1.5,0) to[out=90,in=180] (0,0.75) to [out=0,in=90] (1.5,0) to[out=-90,in=0] (0,-0.75);

            \begin{scope}[scale=0.8]
                \filldraw[fill=teal!40] (0,-0.75) to[out=180,in=-90] (-1.5,0) to[out=90,in=180] (0,0.75) to [out=0,in=90] (1.5,0) to[out=-90,in=0] (0,-0.75);
                \node (Qslefttop) at (-1.5,0) {};
                \node (Qsrighttop) at (1.5,0) {};
            \end{scope}

            \begin{scope}[scale=0.5, xshift=-0.6cm]
                \filldraw[fill=orange!40] (0,-0.75) to[out=180,in=-90] (-1.5,0) to[out=90,in=180] (0,0.75) to [out=0,in=90] (1.5,0) to[out=-90,in=0] (0,-0.75);
                \node (Qstablefttop) at (-1.5,0) {};
                \node (Qstabrighttop) at (1.5,0) {};
            \end{scope}

            \fill[purple] (-0.7,0) node (ta) {} circle[radius=0.05cm];
            \fill[purple] (0.3,0.4) node (tb) {} circle[radius=0.05cm];
            \fill[violet] (0.6,-0.3) node (tc) {} circle[radius=0.05cm];
            \fill[violet] (-0.1,-0.1) node (td) {} circle[radius=0.05cm];
        \end{scope}
        \end{scope}

        \begin{scope}[yshift=-0.5cm]
            \begin{scope}[scale=1.3]
                \filldraw[fill=lightgray!20] (0,-0.75) to[out=180,in=-90] (-1.5,0) to[out=90,in=180] (0,0.75) to [out=0,in=90] (1.5,0) to[out=-90,in=0] (0,-0.75);

            \begin{scope}[scale=0.8]
                \filldraw[fill=teal!40] (0,-0.75) to[out=180,in=-90] (-1.5,0) to[out=90,in=180] (0,0.75) to [out=0,in=90] (1.5,0) to[out=-90,in=0] (0,-0.75);
                \node (Cclefttop) at (-1.5,0) {};
                \node (Ccrighttop) at (1.5,0) {};
            \end{scope}

            \begin{scope}[scale=0.5, xshift=-0.6cm]
                \filldraw[fill=orange!40] (0,-0.75) to[out=180,in=-90] (-1.5,0) to[out=90,in=180] (0,0.75) to [out=0,in=90] (1.5,0) to[out=-90,in=0] (0,-0.75);
            \end{scope}

            \fill[purple] (-0.7,0) node (ma) {} circle[radius=0.05cm];
            \fill[purple] (0.3,0.4) node (mb) {} circle[radius=0.05cm];
            \fill[violet] (0.6,-0.3) node (mc) {} circle[radius=0.05cm];
            \fill[violet] (-0.1,-0.1) node (md) {} circle[radius=0.05cm];
        \end{scope}
        \end{scope}
        
        \begin{scope}[yshift=-3.5cm]
            \begin{scope}[scale=1.3]
                \filldraw[fill=lightgray!20] (0,-0.75) to[out=180,in=-90] (-1.5,0) to[out=90,in=180] (0,0.75) to [out=0,in=90] (1.5,0) to[out=-90,in=0] (0,-0.75);

            \begin{scope}[scale=0.8]
                \filldraw[fill=teal!40] (0,-0.75) to[out=180,in=-90] (-1.5,0) to[out=90,in=180] (0,0.75) to [out=0,in=90] (1.5,0) to[out=-90,in=0] (0,-0.75);
                \node (Narainlefttop) at (-1.5,0) {};
                \node (Narainrighttop) at (1.5,0) {};
            \end{scope}

            \begin{scope}[scale=0.5, xshift=-0.6cm]
                \filldraw[fill=orange!40] (0,-0.75) to[out=180,in=-90] (-1.5,0) to[out=90,in=180] (0,0.75) to [out=0,in=90] (1.5,0) to[out=-90,in=0] (0,-0.75);
                \node (Nstablefttop) at (-1.5,0) {};
                \node (Nstabrighttop) at (1.5,0) {};
            \end{scope}

            \fill[purple] (-0.7,0) node (ba) {} circle[radius=0.05cm];
            \fill[purple] (0.3,0.4) node (bb) {} circle[radius=0.05cm];
            \fill[violet] (0.6,-0.3) node (bc) {} circle[radius=0.05cm];
            \fill[violet] (-0.1,-0.1) node (bd) {} circle[radius=0.05cm];
        \end{scope}
        \end{scope}
        
        \begin{scope}[xshift=-5.0cm, yshift=-3.5cm]
            \begin{scope}[scale=1.3]
                \filldraw[fill=lightgray!20] (0,-0.75) to[out=180,in=-90] (-1.5,0) to[out=90,in=180] (0,0.75) to [out=0,in=90] (1.5,0) to[out=-90,in=0] (0,-0.75);

            \begin{scope}[scale=0.8]
                \filldraw[fill=teal!40] (0,-0.75) to[out=180,in=-90] (-1.5,0) to[out=90,in=180] (0,0.75) to [out=0,in=90] (1.5,0) to[out=-90,in=0] (0,-0.75);
                \node (Fermilefttop) at (-1.5,0) {};
                \node (Fermirighttop) at (1.5,0) {};
            \end{scope}

            \fill[purple] (-0.7,0) node (bla) {} circle[radius=0.05cm];
            \fill[purple] (0.3,0.4) node (blb) {} circle[radius=0.05cm];
            \fill[violet] (0.6,-0.3) node (blc) {} circle[radius=0.05cm];
            \fill[violet] (-0.1,-0.1) node (bld) {} circle[radius=0.05cm];
        \end{scope}
        \end{scope}

        \begin{scope}[xshift=5.0cm, yshift=-3.5cm]
            \begin{scope}[scale=1.3]
                \filldraw[fill=lightgray!20] (0,-0.75) to[out=180,in=-90] (-1.5,0) to[out=90,in=180] (0,0.75) to [out=0,in=90] (1.5,0) to[out=-90,in=0] (0,-0.75);

            \begin{scope}[scale=0.8]
                \filldraw[fill=teal!40] (0,-0.75) to[out=180,in=-90] (-1.5,0) to[out=90,in=180] (0,0.75) to [out=0,in=90] (1.5,0) to[out=-90,in=0] (0,-0.75);
                 \node (Orbilefttop) at (-1.5,0) {};
                \node (Orbirighttop) at (1.5,0) {};
            \end{scope}

            \fill[purple] (-0.7,0) node (bra) {} circle[radius=0.05cm];
            \fill[purple] (0.3,0.4) node (brb) {} circle[radius=0.05cm];
            \fill[violet] (0.6,-0.3) node (brc) {} circle[radius=0.05cm];
            \fill[violet] (-0.1,-0.1) node (brd) {} circle[radius=0.05cm];
        \end{scope}
        \end{scope}

        \begin{scope}[decoration={markings, mark=at position 0.5 with {\arrow{>}}}]
        \end{scope}

        \draw[->, >=stealth, purple, thick] (-5.3,-2.1) to (bla);
        \draw[->, >=stealth, purple, thick] (-5.3,-2.1) to (blb);

        \draw[->, >=stealth, violet, thick] (5.3,-2.1) to (brc);
        \draw[->, >=stealth, violet, thick] (5.3,-2.1) to (brd);

        \draw[->, >=stealth, orange, thick] (3.5,1.3) to (Qstabrighttop);

        \draw[thick, purple, font=\small] (-5.3,-1.8) node {\textrm{SUSY CFTs}};
        \draw[thick, violet, font=\small] (5.3,-1.8) node {\textrm{Self-dual CFTs}};
        \draw[thick, orange, font=\small] (3.7,1.0) node {\textrm{Qudit stabilizer codes}};
        \draw[thick] (-5,2.5) node {\textrm{Qudit subsystem codes}};
        \draw[thick] (-5,-0.5) node {\textrm{Classical codes}};
        \draw[thick] (0,-5) node {\textrm{Narain CFTs}};
        \draw[thick] (5,-5) node {\textrm{Orbifolded CFTs}};
        \draw[thick] (-5,-5) node {\textrm{Fermionized CFTs}};

        \begin{scope}[decoration={markings, mark=at position 0.5 with {\arrow{>}}}]
            \draw[dashed, postaction=decorate] (Cclefttop)--(Narainlefttop);
            \draw[dashed, postaction=decorate] (Ccrighttop)--(Narainrighttop);
            \draw[dashed, postaction=decorate] (Qslefttop)--(Cclefttop);
            \draw[dashed, postaction=decorate] (Qsrighttop)--(Ccrighttop);
            \draw[dashed, postaction=decorate, decoration={markings, mark=at position 0.25 with {\arrow{>}}}, decoration={markings, mark=at position 0.75 with {\arrow{>}}}] (Qstablefttop)--(Nstablefttop);
            \draw[dashed, postaction=decorate, decoration={markings, mark=at position 0.25 with {\arrow{>}}}, decoration={markings, mark=at position 0.75 with {\arrow{>}}}] (Qstabrighttop)--(Nstabrighttop);
            \draw[dashed, postaction=decorate] (Narainlefttop)--(Fermirighttop);
            \draw[dashed, postaction=decorate] (Narainrighttop)--(Orbilefttop);
        \end{scope}

        \draw[thick] (2.1,-2.0) node {\textrm{\scriptsize{\cite{Yahagi:2022idq}}}};
         \draw[thick] (0,1) node {\textrm{\scriptsize{\cite{Dymarsky:2020qom,Kawabata:2022jxt}}}};
         \draw[thick] (2.5,-4.0) node {\textrm{\scriptsize{\cite{Kawabata:2023usr,Kawabata:2023iss}}}};
         \draw[thick] (-2.5,-4.0) node {\textrm{\scriptsize{\cite{Kawabata:2023usr,Kawabata:2023iss}}}};
        \end{tikzpicture}
    \end{center}
        \caption{
    An illustration for our construction of Narain CFTs from qudit subsystem codes. 
    Narain CFTs have been constructed from a certain class of quantum error-correcting codes called stabilizer codes~\cite{Dymarsky:2020qom,Kawabata:2022jxt} (the orange ellipses), which is a subset of subsystem codes. 
    We find the new construction of Narain CFTs by extending stabilizer codes to subsystem codes (the light green ellipses). In this construction, we utilize the existence of a map from subsystem codes to classical codes and the fact that Narain code CFTs can be built out of a certain class of classical codes~\cite{Yahagi:2022idq}. 
    We also investigate their orbifold and fermionization by gauging global $\BZ_2$ symmetries following~\cite{Kawabata:2023usr,Kawabata:2023iss}.
    We find a new discrete set of bosonic CFTs that are self-dual under gauging (purple points) and supersymmetric CFTs (red points).
   }
    \label{fig:Const_Narain_code}
    \end{figure}

Our construction of code CFTs expands the previous one for stabilizer codes over finite field $\BF_p$ with prime $p$ \cite{Dymarsky:2020qom,Kawabata:2022jxt} to subsystem codes over $\BF_p$, and proceeds as follows.
First, we map a quantum subsystem $[[n,k,r]]_p$ code to the corresponding classical linear $[2n, n + r - k]_p$ code by representing the gauge group $\CG$ of the former as the generator matrix of the latter.
We then produce a $2n$-dimensional lattice by uplifting the codewords to lattice vectors by the so-called Construction A \cite{conway2013sphere}.
Finally, the Narain code CFT can be constructed by identifying the resulting lattice as the momentum lattice of a Narain CFT.
While our construction can be applied to any subsystem code, the resulting CFT is not necessarily modular invariant.
The code CFTs constructed from classical codes are examined in \cite{Yahagi:2022idq} and the condition for the resulting CFTs being modular invariant is obtained there.
We adapt the results of \cite{Yahagi:2022idq} to our case and find that the modular invariant code CFTs can be constructed from quantum subsystem $[[n,k,k]]_p$ codes satisfying the condition \eqref{CFT_condition}.
This condition is less restrictive than the one for stabilizer codes in \cite{Dymarsky:2020qom,Kawabata:2022jxt}, which allows us to explore a broader class of code CFTs than before. (See figure \ref{fig:Const_Narain_code} for the summary of our construction.)
Indeed, we find a new class of subsystem codes satisfying the condition \eqref{CFT_condition}, called the B-form codes, that cannot be realized as stabilizer codes.

All code CFTs constructed in the above manner have a property independent of the code structure such that they possess at least one global $\BZ_2$ symmetry.
Hence, the orbifolded and fermionized theories of the code CFTs can be defined by gauging the $\BZ_2$-symmetry if it is non-anomalous \cite{Kawabata:2023usr,Kawabata:2023iss}.
The fermionized theories of the code CFTs have been examined in the previous studies and a few stabilizer codes that yield supersymmetric CFTs have been discovered \cite{Kawabata:2023usr,Kawabata:2023iss}.
We extend this program to subsystem codes and exploit their fermionic code CFTs to search for new examples of supersymmetric code CFTs.
The necessary conditions for fermionic CFTs being supersymmetric are given in \cite{Bae:2021jkc,Bae:2021lvk}, which can be tested through their torus partition functions that encode the operator spectrum.
For code CFTs, one can represent the torus partition function by using the complete weight enumerator polynomials of the associated codes, and it is straightforward to check whether the necessary conditions are met for a given fermionic code CFT.

Focusing on the B-form subsystem codes over $\BF_p$, we show that the code CFTs can be classified by oriented graphs with weights on the edges.
For $p=2$, our classification simplifies to that of code CFTs employing unoriented graphs in \cite{Dymarsky:2020qom}.
We leverage the classification of both unoriented and oriented graphs to enumerate the B-form code CFTs over $\BF_2$ and $\BF_3$ with small central charges, respectively.
To investigate the properties of the B-form code CFTs, we calculate the torus partition functions as well as those of their orbifolded and fermionized theories when the $\BZ_2$-gauging is possible.
From the fermionic partition functions, we identify several new fermionic code CFTs that meet the requisite criteria for being supersymmetric.
For some of those that have non-vanishing Witten indices, we prove that there exist supercurrents that are realized as a linear combination of vertex operators.
On the other hand, the rest of them have vanishing Witten indices and hence may have spontaneously broken supersymmetry.

For the B-form code CFTs, we compare the torus partition functions and the orbifold ones and find a few examples that are self-dual under the $\BZ_2$-orbifold.
In particular, code CFTs whose fermionized theories have vanishing Witten indices are always self-dual under the $\BZ_2$ gauging.
In recent studies, theories self-dual under gauging have attracted special attention in the context of non-invertible symmetries (see e.g., \cite{Shao:2023gho} for a review). 
By applying the half gauging technique in \cite{Choi:2021kmx}, one can construct non-invertible duality defects from self-dual theories.
We expect that our constructions of the code CFTs and their gaugings are beneficial in searching for CFTs with non-invertible symmetries.

This paper is organized as follows. In section \ref{ss:review_subsystem}, we start to review the quantum subsystem codes for qudit systems. 
We elucidate the structure and properties of subsystem codes with an emphasis on the relationship between subsystem codes and classical codes through the symplectic representation of the (generalized) Pauli operators.
In section \ref{ss:subsystem_cft}, we describe the construction of Narain CFTs from subsystem codes via Lorentzian self-dual lattices.
We show that they have global $\BZ_2$ symmetries and identify when they are non-anomalous.
We then construct the orbifolded and fermionized theories of the code CFTs by gauging the non-anomalous $\BZ_2$ symmetries, and represent their torus partition functions using the complete weight enumerator polynomials of the associated codes.
In section \ref{ss:fermion}, we discuss supersymmetry that fermionized code CFTs could have and the realization of supercurrents as vertex operators.
In section \ref{ss:enumeration}, we investigate and enumerate the examples of the B-form code CFTs with small central charges by classifying the weighted oriented graphs ($p$ is odd prime) and unoriented graphs ($p=2$).
By gauging non-anomalous global $\BZ_2$ symmetries, we find a discrete set of supersymmetric CFTs and bosonic CFTs self-dual under the $\BZ_2$ symmetries that have not been constructed from quantum codes before.
Finally, section \ref{ss:discussion} concludes with discussions and future directions.

\section{Quantum subsystem codes}\label{ss:review_subsystem}

In this section, we review a class of quantum error-correcting codes called quantum subsystem codes. This framework is wider than well-known stabilizer codes and introduces a new ingredient called gauge qudits. In section~\ref{ss:qudit}, we introduce a qudit system with prime dimensions and operators acting on it.
Section~\ref{ss:subsystem} is devoted to giving a simple example of subsystem codes to illustrate the concept of gauge qubits and the error correction procedure in subsystem codes.
Finally, we give an operator formalism of subsystem codes by using a group-theoretical method in section~\ref{ss:stabilizer}.

\subsection{Qudit system}
\label{ss:qudit}
We consider a $p$-ary qudit system whose Hilbert space $\CH_p$ is spanned by an orthonormal basis $\{\, \ket{0}, \ket{1}, \cdots, \ket{p-1}\,\}$.
For a prime $p$, $\BF_p \equiv \BZ/p\,\BZ$ becomes a finite field.
In what follows, we will focus on this case for simplicity.
The fundamental Pauli operators $X$ and $Z$ acting on a single qudit system are defined by~\cite{Gottesman:1998se}
\begin{align}
    X\,\ket{x} = \ket{x+1}\ , \qquad \quad Z\,\ket{x} = \omega_p^x\,\ket{x} \ ,
\end{align}
where $\omega_p = \exp(2\pi \i/p)$.
More concretely, we can write these operators as
\begin{align}
    X = \sum_{x\in\BF_p} \ket{x+1}\bra{x}\,,\qquad
    Z = \sum_{x\in\BF_p} \omega_p^x\ket{x}\bra{x}\,.
\end{align}
Then, we can read off the commutation relation $XZ = \omega_p^{-1}\, ZX$.

There is a subtle difference in the global phase of the Pauli operators between $p=2$ and other cases.
An element in the Pauli group acting on a single qudit system is given by $\mathsf{g}(a,b)=\omega^\kappa X^{a}Z^b$ where the phase factor is
\begin{align}
    \omega = 
        \begin{dcases}
            \i & (p=2) \ , \\
            \omega_p & (p:\text{odd prime})\ ,
        \end{dcases} 
\end{align}
and $\kappa\in\BZ_4$ for $p=2$ and $\kappa\in\BF_p$ for an odd prime $p$.
This is because, from the commutation relation between $X$ and $Z$, we have
\begin{align}
    (XZ)^p = 
    \begin{dcases}
        -1 & (p=2)\,,\\
        +1 & (p:\text{odd prime})\,.
    \end{dcases}
\end{align}
In the binary case $(p=2)$, we need the imaginary number $\i$ to define the Pauli Y operator $Y = \i X Z$ to hold the condition $Y^2=1$.
On the other hand, for an odd prime $p$, the generalized Pauli Y-like operator can be defined simply by $Y =X^aZ^b$ $(a,b\in\BF_p)$, which automatically satisfies $Y^p=1$.

We can write the generalized Pauli operators on an $n$-qudit system by taking the $n$-fold tensor product.
For a pair of $n$-vectors $\Ba = (a_1, \cdots, a_n), \Bb = (b_1, \cdots, b_n)$ in $\BF_p^n$, we define an operator $g(\Ba, \Bb)$ acting on an $n$-qudit system by
\begin{align}
    g(\Ba, \Bb) \equiv X^{a_1} Z^{b_1}\otimes \cdots \otimes X^{a_n} Z^{b_n}\ .
\end{align}
The generalized Pauli group $\CP_n$ consists of an element $\omega^\kappa\,g(\Ba, \Bb)$ where $\kappa\in\BZ_4$ for $p=2$ and $\kappa\in\BF_p$ for an odd prime $p$.
A pair of two operators satisfy the relation
\begin{align}
\label{eq:com_rel}
    g(\Ba, \Bb) \, g(\Ba', \Bb')  = \omega_p^{-\Ba\cdot \Bb' + \Ba'\cdot \Bb}\,g(\Ba', \Bb') \, g(\Ba, \Bb) \ ,
\end{align}
where $\Ba\cdot\Bb = \sum_{i=1}^n a_i\,b_i$.
The generalized Pauli group $\CP_n$ can be mapped to the vector space of dimension $2n$ over $\BF_p$.
Any operator in the Pauli group can be denoted by $g(\Ba,\Bb)$ up to a phase factor, and thus specified by a vector $(\Ba,\Bb)\in\BF_p^{2n}$ of length $2n$.
To recover the commutation relation of Pauli operators, we introduce a symplectic form
\begin{align}
    \sf{W} = \begin{bmatrix}
        0 & -I_n\\
        I_n & 0
    \end{bmatrix}\,.
\end{align}
Since the commutation relation is given by \eqref{eq:com_rel}, two Pauli operators $g(\Ba,\Bb)$ and $g(\Ba',\Bb')$ commute if and only if $(\Ba,\Bb)\circ (\Ba',\Bb') = (\Ba,\Bb)\, \sf{W} \,(\Ba',\Bb')^T  = 0\in\BF_p$ where $\circ$ denotes the product by the symplectic form $\sf{W}$.

\subsection{Subsystem codes}
\label{ss:subsystem}

We describe the concept of subsystem codes using a simple example.
Subsystem codes are a wide class of quantum error-correcting codes, which include stabilizer codes.
This framework leads to more efficient ways of error correction with less error syndrome by adding a new degree of freedom called gauge qubits.
For example, the subsystem version of Shor's nine qubit code can correct the same errors as Shor's nine qubit stabilizer code with fewer measurements, which leads to a simpler decoding procedure \cite{poulin2005stabilizer}.

Throughout this subsection, we focus on the binary case $(p=2)$ for simplicity. The generalization to qudits is straightforward.

Consider an $n$-qubit system and divide $n$ qubits into three parts: the first $s$ and the second $r$ qubits in an imperfect noisy environment, and the third $k$ qubits in a perfect noiseless environment where $s+r+k=n$.
A state of the $n$-qubit system can be represented as
\begin{align}
     \ket{a_1\dots a_s}\otimes \ket{a_{s+1}\dots a_{s+r}}\otimes \ket{a_{s+r+1}\dots a_n}\,,
\end{align}
where $a_i\in\BF_2$.
Suppose the information is encoded into the last $k$ qubits called logical qubits.
Let us initialize the first $s+r$ qubits by $\ket{0^{s+r}}$ and we perform unitary transformations on the last $k$ qubits to manipulate the information. Ideally, we obtain
\begin{align}
\label{eq:desired}
    \ket{\psi} = \ket{0^s}\otimes\ket{0^r}\otimes\ket{\psi_L}\,.
\end{align}
However, after some time, ambient noise corrupts the first $s+k$ qubits
\begin{align}
\label{eq:corrupted}
    \ket{\psi'} = \ket{\psi_S}\otimes\ket{\psi_G}\otimes\ket{\psi_L}\,.
\end{align}
Generally, quantum error-correcting codes give a recovery map from the corrupted state~\eqref{eq:corrupted} to the desired state~\eqref{eq:desired}.
Subsystem codes also give the recovery map but neglect the difference of states in the second $r$ qubits called gauge qubits.

We can recover the corrupted states up to gauge qubits as follows.
First, measure the first $s$ qubits by $Z_i$ $(i=1,2,\cdots,s)$ where $Z_i$ is the Pauli Z matrix acting on the $i$-th qubit, which projects the corrupted state to
\begin{align}
    \ket{\psi''} = \ket{m_1\dots m_s}\otimes\ket{\psi_G}\otimes\ket{\psi_L}\,,
\end{align}
where $(m_1,\cdots,m_s)$ is a set of outcomes taking values $\pm1$.
Then, we apply the recovery map
\begin{align}
    R = X^{m_1}\otimes \dots\otimes X^{m_s} \otimes I_{2^{r+k}\times2^{r+k}}\,,
\end{align}
and we finally obtain the corrected state up to gauge qubits
\begin{align}
    \ket{\widetilde{\psi}\,} = \ket{0^s}\otimes\ket{\psi_G}\otimes\ket{\psi_L}\,.
\end{align}

In this example, the code subspace $\CH_C$, which we protect from the noise, consists of states represented by 
$\ket{0^s}\otimes\ket{\psi_G}\otimes\ket{\psi_L}$ where $\ket{\psi_G}\in (\BC^{2})^{\otimes r}$ and $\ket{\psi_L}\in (\BC^{2})^{\otimes k}$.
They are defined by the subspace invariant under the action of $S_i = Z_i$ $(i=1,2,\cdots,s)$.
These Pauli operators generate an abelian group $\CS=\langle I,S_1,\dots,S_s\rangle$ called the stabilizer group. Thus, 
\begin{align}
    \CH_C = \left\{\; \ket{\psi}\in (\BC^2)^{\otimes n}\;\middle|\; S_i\ket{\psi} = \ket{\psi},\; S_i\in \CS\; \right\}\,.
\end{align}
Note that each generator $g_i$ projects the whole Hilbert space $\CH$ into half and finally we have
$\CH = \CH_C \oplus \CH_C^\perp$ where $\CH_C\cong (\BC^2)^{\otimes (n-s)}=(\BC^2)^{\otimes (r+k)}$.

The code subspace $\CH_C$ has the subsystem structure $\CH_C = \CH_G\otimes \CH_L$ where $\CH_G$ denotes a subsystem spanned by gauge qubits and $\CH_L$ denotes one by logical qubits. Correspondingly,
\begin{align}
    \CH = \CH_G\otimes \CH_L \oplus \CH_C^\perp\,.
\end{align}
The Pauli operators $X_{s+r+j}$, $Z_{s+r+j}$ ($j=1,2,\cdots,k$) acting on subsystem $\CH_L$ are called the logical operators.
On the other hand, the Pauli operators $X_{s+j}$, $Z_{s+j}$ acting on subsystem $\CH_G$ generate the Pauli group of $r$ qubits. 
Together with stabilizer group $\CS$ and $\i\, I$, the set of operators form a group called the gauge group
\begin{align}
    \CG = \langle \, \i\, I , S_1,\dots,S_s, X_{s+1},Z_{s+1},\dots,X_{s+r},Z_{s+r}\,\rangle\,,
\end{align}
where the bracket $\langle\cdot\rangle$ denotes a generating set of the group.
This group leaves the encoded information invariant under the action.
For a non-zero gauge qubit, the gauge group is non-abelian.
Note that an ordinary stabilizer code is a subsystem code without gauge qubits. In this case, the gauge group is abelian.

The above example is trivial because the logical qubits are isolated physically.
To consider a non-trivial case, we apply a unitary transformation (Clifford transformation) to the above system.
Then, the fundamental Pauli operators $X_i'=U^\dagger \,X_i\, U$, $Z_i' = U^\dagger\, Z_i\, U$ are no longer local operators acting on a single qubit.
Rather they globally act on the $n$-qubit system while they have the same commutation relations
\begin{align}
    X_i'\, Z_j' = (-1)^{\delta_{ij}} \,Z_j' \,X_i'\,.
\end{align}
These operators $X_i'$, $Z_i'$ generate the generalized Pauli group $\CP_n$ and behave as Pauli operators acting on a single qubit.
However, the single qubit is a virtual qubit rather than a bare qubit related to the original Pauli operators $X_i$, $Z_i$.
For the virtual qubits, we can analogously construct subsystem codes.
Without loss of generality, we can take a stabilizer group by $\CS = \langle S_1,\dots,S_s\rangle$ where $S_j = Z_j'$ and $s\leq n$ and can take a gauge group by
\begin{align}
    \CG = \langle \, \i\, I , S_1,\dots,S_s, X_{s+1}',Z_{s+1}',\dots,X_{s+r}',Z_{s+r}'\,\rangle\,,
\end{align}
where $s+r\leq n$.
The only difference with the trivial example is whether the operator is primed or unprimed.
To explain it in detail, we give a sophisticated tool for defining subsystem codes by an operator formalism in the next subsection.

\subsection{Stabilizer formalism of subsystem codes}
\label{ss:stabilizer}

We consider an $n$-qudit system and establish the stabilizer formalism for subsystem codes.
Let $X_j',Z_j'$ be the Pauli operators acting on the $j$-th virtual qudit, which satisfy the commutation relation
\begin{align}
    X_i'\, Z_j' = \omega_p^{-\delta_{ij}} \,Z_j' \,X_i'\,.
\end{align}
Without loss of generality, the stabilizer group can be chosen as
\begin{align}
    \CS = \langle \, Z_1',Z_2',\dots,Z_s' \,\rangle\,,
\end{align}
since the generators $Z_j'$ are commuting with each other. We denote the stabilizer generator as $S_j = Z_j'$.
The stabilizer group defines the code subspace of $p^{n-s}$ dimensions:
\begin{align}
\label{code_space}
    \CH_C = \left\{\, \ket{\psi}\in (\BC^p)^{\otimes n}\,\, \big|\,\, g\,\ket{\psi} = \ket{\psi},\, ~\forall g \in \CS \,\right\} \ .
\end{align}
The set of Pauli operators that map the code subspace into itself is the normalizer $N(\CS)$ of stabilizer group $\CS$ in the generalized Pauli group $\CP_n$
\begin{align}
    N(\CS) =  \langle \, \omega I , S_1,\dots,S_s, X_{s+1}',Z_{s+1}',\dots,X_{n}',Z_{n}'\,\rangle\,.
\end{align}
The normalizer $N(\CS)$ consists of two sets of Pauli operators: the gauge group $\CG$ and the group $\CL = N(\CS)/\CG$ generated by logical operators.
We define the gauge group $\CG$ by the set of operators that do not change encoded information.
Since $\langle \omega I\rangle$ and the stabilizer group $\CS$ do not affect states in $\CH_C$, they should be in the gauge group $\CG$.
Additionally, to include the subsystem structure $\CH_C = \CH_G\otimes \CH_L$ on the code subspace, we need $[\CG,\CL]=0$~\cite{Zanardi:2004zz}.
Correspondingly, we can always take the gauge group by
\begin{align}
    \CG =  \langle \, \omega I , S_1,\dots,S_s, X_{s+1}',Z_{s+1}',\dots,X_{s+r}',Z_{s+r}'\,\rangle\,,
\end{align}
where $s+r\leq n$ and the group $\CL$ of logical operators by $\CL = \langle\, X_{s+r+1}',Z_{s+r+1}',\dots, X_{n}',Z_{n}' \,\rangle$ where we set $s+r+k = n$.
On the code subspace $\CH_C$, a gauge operator $g\in\CG$ and logical operator $L\in\CL$ act by
\begin{align}
    g = g^A\otimes {\bf{1}}^B_{2^k}\,,\qquad L = {\bf 1}^A_{2^r}\otimes L^B\,,
\end{align}
for some Pauli operators $g^A$, $L^B$ acting on $\CH_G$ and $\CH_L$, respectively.
Thus, the code subspace has the subsystem structure $\CH_C = \CH_G\otimes \CH_L$ where $\CH_G\cong (\BC^p)^r$ and $\CH_L\cong (\BC^p)^k$.
The subsystem code is called an $[[n,k,r]]_p$ type, which has the code subspace consisting of $k$ logical qudits and $r$ gauge qudits in an $n$-qudit system.

Conversely, a given gauge group $\CG\subset\CP_n$ determines a subsystem code.\footnote{Following the convention of~\cite{poulin2005stabilizer}, we include a phase factor $\omega I\in \CP_n$ as a generator in a gauge group $\CG$.} 
The center of the gauge group defines the stabilizer group by $\CS = \CZ(\CG)/\langle \omega I\rangle = \langle\, S_1, \cdots, S_{s}\,\rangle$, which is generated by a commuting set of operators $S_i = g\left(\Ba^{(i)}, \Bb^{(i)}\right)~(i=1, \cdots, s)$.
The remaining generators of $\CG$ can be chosen as the Pauli operators $ X_{s+i}', Z_{s+i}'~(i=1,\cdots, r)$ which act on the $r$ gauge qudits and satisfy
\begin{align}
    X_{s+i}'\, Z_{s+j}' = \omega_p^{-\delta_{ij}}\, Z_{s+j}'\, X_{s+i}' \ .
\end{align}
They generate, together with the stabilizer group, the whole gauge group $\CG$ as
\begin{align}\label{subsystem_standard}
    \CG =  \langle \, \omega I , S_1,\dots,S_s, X_{s+1}',Z_{s+1}',\dots,X_{s+r}',Z_{s+r}'\,\rangle\,.
\end{align}
Since these operators are elements of the generalized Pauli group, there exist $r$ pairs of vectors $\left(\Ba^{(j)}, \Bb^{(j)}\right)$ such that $ X_{s+i}' \propto g\left(\Ba^{(s+2i-1)}, \Bb^{(s+2i-1)}\right), Z_{s+i}' \propto g\left(\Ba^{(s+2i)}, \Bb^{(s+2i)}\right)~(i=1,\cdots, r)$.
Then, the gauge group $\CG$ can be represented as an $(s+2r) \times 2n$ matrix $\SH_\text{sub}$ over $\BF_p$:
\begin{align}\label{subsystem_H}
    \SH_\text{sub} = 
    \left[
        \begin{array}{c|c}
    	~\Ba^{(1)} ~& ~\Bb^{(1)}~ \\
            ~\vdots ~& ~\vdots~ \\
        ~\Ba^{(s)} ~& ~\Bb^{(s)}~ \\ \hline
        ~\Ba^{(s+1)} ~& ~\Bb^{(s+1)}~ \\
            ~\vdots ~& ~\vdots~ \\
            ~\Ba^{(s+2r)}~ & ~\Bb^{(s+2r)} ~
        \end{array}
    \right] \ .
\end{align}
A subsystem code with $\SH_\text{sub}$ of the standard form \eqref{subsystem_H} satisfies the condition:
\begin{align}\label{subsystem_condition}
    \SH_\text{sub}\,\SW\,\SH_\text{sub}^T 
        = 
        \left[
        \begin{array}{c|ccccc}
        ~O_{s,s} ~& & & ~O_{s,2r}~ & & \\ \hline
            &  ~0~ & 1 & & & \\
            & -1 &  ~0~ & & & \\
        ~O_{2r,s} ~&  & & \ddots & & \\
            &  & & &~0~ & 1  \\
            &  & & & -1 &  ~0~              
        \end{array}
    \right]         
        \quad \text{mod}~p \ ,
\end{align}
where $O_{m,n}$ is the $m \times n$ zero matrix.
Note that we can remove $s$ by using the relation $n=s+k+r$ as $s+2r = n+r - k$, so $\SH_\text{sub}$ is an $(n+r-k)\times 2n$ matrix.

\paragraph{Example: $[[3,1,1]]_2$ subsystem code.}
We consider a $3$-qubit system.
Let us see the gauge group $\CG$ generated by
\begin{align}
        \CG=\langle\, \i\,I,\,ZZI,\,XXI,\,IZZ\,\rangle.
\end{align}
Then, the stabilizer group is given by its center $\CS = \CZ(\CG)/\langle\i I\rangle = \langle Z_1' = ZZI\rangle$. The other non-trivial generators are the Pauli operators acting on a virtual gauge qubit: $X_2'=XXI$, $Z_2'=IZZ$. Therefore, the gauge group can be written as
\begin{align}
     \CG=\langle \, \i\,I,\,Z_1',\,{X}'_2,\,{Z}'_2\, \rangle \ .
\end{align}
In this setup, the other generators of the Pauli group are determined by the commutation relation
\begin{align}
    X_1' = IXX\,,\quad X_3' = XXX\,,\quad Z_3' = ZZZ\,.
\end{align}
Using these symbols, the Pauli group can be written as $\CP_3=\langle \i I, X_1',Z_1',X_2',Z_2',X_3',Z_3'\rangle$.
Then, the normalizer $N(\CS)$ of the stabilizer group in $\CP_3$ is given by 
\begin{align}
    N(\CS) = \langle\, \i\, I, \,Z_1',\, X_2',\, Z_2', \,X_3',\,Z_3'\,\rangle\,.
\end{align}
The logical operators in $\CL = N(\CS)/\CG$ are specified by
\begin{align}
    \CL = \langle X_3'=XXX\,, Z_3'=ZZZ\rangle\,.
\end{align}

The code subspace $\CH_C$ is the invariant subspace under the action of $Z_1'=ZZI$
\begin{align}
    \CH_C=\text{Span}_{\BC}\left\{\ket{000},\,\ket{001},\,\ket{110},\,\ket{111}\right\}\,.
\end{align}
The correspondence between the representation by virtual qubits and by bare qubits is
\begin{equation}
        \begin{aligned}
       &\ket{\Tilde{0},\Tilde{0}}_{G}\otimes\ket{\Tilde{0}}_L=\ket{000}\,,\ \ 
        \ket{\Tilde{0},\Tilde{1}}_{G}\otimes\ket{\Tilde{0}}_L=\ket{110}\,,\\
        &\ket{\Tilde{0},\Tilde{0}}_{G}\otimes\ket{\Tilde{1}}_L=\ket{111}\,,\ \ 
        \ket{\Tilde{0},\Tilde{1}}_{G}\otimes\ket{\Tilde{1}}_L=\ket{001}\,.
   \end{aligned}
\end{equation}
Thus, the code subspace $\CH_C$ has the subsystem structure
\begin{align}
\CH_C=\left(\alpha\ket{\Tilde{0},\Tilde{0}}_{G}+\beta\ket{\Tilde{0},\Tilde{1}}_{G}\right)\otimes\left(\gamma\ket{\Tilde{0}}_L+\delta\ket{\Tilde{1}}_L\right),
\end{align}
where $\alpha,\beta,\gamma,\delta\in\BC$.
We can see that Pauli operators in the gauge group $\CG$ act on only first and second virtual qubits.
The generator matrix corresponding to this gauge group $\CG$ is
   \begin{align}
      \SH_{\text{sub}}=\begin{bmatrix}
            g^{-1}({Z}_{1}')\\
            g^{-1}({X}_{2}')\\
            g^{-1}({Z}_{2}')
        \end{bmatrix}
        =\begin{bmatrix}
           0&0&0&1&1&0\\
           1&1&0&0&0&0\\
           0&0&0&0&1&1
        \end{bmatrix}.
  \end{align}
  
\paragraph{Relation with classical codes.}

We associate an $[[n,k,r]]_p$ subsystem code whose gauge group is specified by $\SH_\text{sub}$ with a classical $[2n, n+r-k]_p$ code $\CC_\text{sub}$ over $\BF_p^{2n}$ with generator matrix $G = \SH_\text{sub}$:
\begin{align}\label{eq:CC-stabilizer}
    \CC_\text{sub} 
        =
        \left\{\, c\in \BF_p^{2n}\,\big|\, c = x\,\SH_\text{sub}, ~x \in \BF_p^{n+r-k}\,\right\} \ .
\end{align}
Conversely, we can construct a subsystem code on an $n$-qudit system from a $[2n,k]_p$ classical code as follows. First let us give the generator matrix of a $[2n,k]_p$ classical code:
\begin{align}
    G = 
    \left[
        \begin{array}{c|c}
    	~\Ba^{(1)} ~& ~\Bb^{(1)}~ \\
            ~\vdots ~& ~\vdots~ \\
        ~\Ba^{(k)} ~& ~\Bb^{(k)}~ 
        \end{array}
    \right] \ .
\end{align}
Then we can define the gauge group of the subsystem code by regarding each row of $\CC$ as the generator of the gauge group:
\begin{align}
    \CG=\left\langle\,\omega I,\,g\left(\Ba^{(1)}, \Bb^{(1)}\right),\,\cdots,\,g\left(\Ba^{(k)}, \Bb^{(k)}\right)\right\rangle.
\end{align}
Thus, the stabilizer group of the subsystem code is determined by the center of the gauge group $\CS = \CZ(\CG)/\langle \omega I\rangle = \langle\, S_1, \cdots, S_{s}\,\rangle\,(s\leq k)$. The generators of the gauge group can be chosen as standard form of the gauge group \eqref{subsystem_standard}. Finally, we get the $[[n,(2n-k-s)/2,(k-s)/2]]_p$ subsystem code and its generator matrix:
\begin{align}
    \SH_\text{sub} = 
    \left[
        \begin{array}{c|c}
    	~\Ba^{(1)} ~& ~\Bb^{(1)}~ \\
            ~\vdots ~& ~\vdots~ \\
        ~\Ba^{(s)} ~& ~\Bb^{(s)}~ \\ \hline
        ~\Ba^{(s+1)} ~& ~\Bb^{(s+1)}~ \\
            ~\vdots ~& ~\vdots~ \\
            ~\Ba^{(k)}~ & ~\Bb^{(k)} ~
        \end{array}
    \right] \ .
\end{align}
Note that this process to find the generators of the standard form of the gauge group is automatically proceeded by performing the Gram-Schmidt process~\cite{haah2013lattice} for the vector space spanned by rows of generator matrix $G$ with the symplectic product defined by
\begin{align}
    (\Ba, \Bb) \circ (\Ba', \Bb')\equiv -\Ba\cdot \Bb' + \Ba'\cdot \Bb,
\end{align}
where $(\Ba, \Bb)\in\BF_p^n\times\BF_p^n$ and $\circ$ denotes the symplectic product.

\paragraph{Example: $[8,4]_3$ Quadratic double circulant code.}
The $[8,4]_3$ quadratic double circulant code is defined by the generator matrix (see appendix \ref{ap:QDC_code} for details):
\begin{align}
    G
            &=\left[
    \begin{array}{cccc|cccc}
        ~1~ & ~0~ & ~0~ & ~0~ & \,~0~ & ~1~ & ~1~ & ~1~\\
        ~0~ & ~1~ & ~0~ & ~0~ & \,-1~ & ~0~ & ~1~ & -1~\\
        ~0~ & ~0~ & ~1~ & ~0~ & \,-1~ & -1~ & ~0~ & ~1~\\
        ~0~ & ~0~ & ~0~ & ~1~ & \,-1~ & ~1~ & -1~ & ~0~
    \end{array}\right]\,.
\end{align}
By reading off the generators of the gauge group from this generator matrix, the gauge group of the subsystem code is
\begin{equation}
    \CG=\left\langle\,\omega_3 I,\,g_1,\,g_2,\,g_3,\,g_4\right\rangle,
\end{equation}
with
\begin{align}
    \begin{aligned}
        g_1&\equiv X\otimes Z\otimes Z\otimes Z\ , & \qquad g_2&\equiv Z^{-1}\otimes X\otimes Z\otimes Z^{-1}\ ,\\
        g_3&\equiv Z^{-1}\otimes Z^{-1}\otimes X\otimes Z\ , & \qquad g_4 &\equiv Z^{-1}\otimes Z\otimes Z^{-1}\otimes X \ ,
    \end{aligned}
\end{align}
where $\omega_3=e^{2\pi \i/3}$ and $X,Z$ are the (generalized) Pauli operators on a qutrit.
To construct the standard form of the gauge group, we rearrange the generators $g_1,\cdots, g_4$ as follows:
\begin{align}
   \begin{aligned}
        Z_{1}' &\equiv \omega_3^{-1}g_1g_2^{-1}g_3=X\otimes X^{-1}\otimes X\otimes I\ , \\
        Z_{2}'&\equiv \omega_3^{-1}g_1^{-1}g_2^{-1}g_4=X^{-1}\otimes X^{-1}\otimes I\otimes X \ ,\\
        X_{3}'&\equiv g_1=X\otimes Z\otimes Z\otimes Z\ , \\
        Z_{3}'&\equiv g_2=Z^{-1}\otimes X\otimes Z\otimes Z^{-1}\ .
   \end{aligned} 
\end{align}
We can easily check that these Pauli operators satisfy commutation relation 
\begin{align}
    X_i'\, Z_j' = \omega_3^{-\delta_{ij}} \,Z_j' \,X_i'\,,
\end{align}
and rewrite the gauge group as:
\begin{align}
     \CG=\langle \, \omega_3 I,\,\CS,\,{X}'_3,\,{Z}'_3\, \rangle \,,
\end{align}
where the stabilizer group is $\CS=\langle\, Z_{1}',\,Z_{2}'\,\rangle$. So we can specify that this subsystem code is $[[4,1,1]]_3$ type and its generator matrix is
\begin{align}
    \SH_\text{sub}=\left[
    \begin{array}{cccc|cccc}
        ~1~ & -1~ & ~1~ & ~0~ & ~0~ & ~0~ & ~0~ & ~0~\\
        -1~ & -1~ & ~0~ & ~1~ & ~0~ & ~0~ & ~0~ & ~0~\\ \hline
        ~1~ & ~0~ & ~0~ & ~0~ & ~0~ & -1~ & ~1~ & ~1~\\
        ~0~ & ~1~ & ~0~ & ~0~ & -1~ & ~0~ & ~1~ & -1~
    \end{array}\right]\,.
\end{align}

\section{Narain CFTs from subsystem codes} 
\label{ss:subsystem_cft}

After reviewing the Narain CFTs and their momentum lattices, we construct the code CFTs from quantum subsystem codes satisfying a certain condition.
We also consider the $\BZ_2$-gauging of the code CFTs to derive the torus partition functions of their orbifolded and fermionized theories when they have non-anomalous $\BZ_2$ symmetries.

\subsection{Review of Narain CFTs}
Narain CFT is a two-dimensional field theory with $n$ compact bosons $X^i(t,\sigma)~ (i=1,\cdots,n)$.
This theory is described by the following action \cite{Narain:1985jj,Narain:1986am}:
\begin{align}
    I
        =
        \dfrac{1}{4\pi\alpha'}\int \d t\int_{0}^{2\pi}\d\sigma\left[\,
        G_{ij}\,(\partial_tX^i\partial_tX^j-\partial_{\sigma}X^i\partial_{\sigma}X^j)
        -
        2B_{ij}\,\partial_tX^i\partial_{\sigma}X^j\,\right],
\end{align}
where $G_{ij}$ is the metric for the target space of the bosons and $B_{ij}$ is the anti-symmetric matrix called the B-field.
The target space of these bosons is compactified on higher dimensional torus with radius $R$:
 \begin{align}
    X^i(t,\sigma)-X^i(t,\sigma+2\pi)=2\pi R\,w^i\ ,
\end{align}
where $w^i\in\BZ\,(i=1,\cdots,n)$ are winding numbers.
This compactification makes the eigenvalue of the left and right momenta discretized as follows:
\begin{align}
    \begin{aligned}
         \hat{p}_{Li}
            &=
            \dfrac{m_i}{R} + \dfrac{R}{\alpha'}\,(B+G)_{ij}\,w^j\ ,\\
         \hat{p}_{Ri}
            &=
            \dfrac{m_i}{R} + \dfrac{R}{\alpha'}\,(B-G)_{ij}\,w^j\ ,
    \end{aligned}
\end{align}
where $m^i\,(i=1,\cdots,n)$ are integers.
We also find it useful to introduce the dimensionless momenta:
\begin{align}
    \begin{aligned}
            p_{L\mu}
                &=
                e^{i}_{\mu}\left[\dfrac{m_i}{r}+\dfrac{r}{2}\,(B+G)_{ij}\,w^j\right] \ ,\\
            p_{R\mu}
                &=
                e^{i}_{\mu}\left[\dfrac{m_i}{r}+\dfrac{r}{2}\,(B-G)_{ij}\,w^j\right]\ ,
    \end{aligned}
\end{align}
where $e^{\mu}_{i}$ are vielbein satisfying $G_{ij}=e^{\mu}_{i}\,e^{\nu}_{j}\,\delta_{\mu\nu}$, $e^{i}_{\mu}$ their inverse, and $r\equiv R\sqrt{2/\alpha^{\prime}}$ is the dimensionless radius.
The left and right momenta construct the momentum lattice $\Tilde{\Lambda}$ whose lattice points are labeled by $(p_L,p_R)\in\BR^{n,n}$.
We can construct another lattice $\Lambda$ by the following coordinate transformation:
\begin{align}
    (\lambda_1,\lambda_2)=\left(\dfrac{p_L-p_R}{\sqrt{2}},\dfrac{p_L+p_R}{\sqrt{2}}\right)\in\Lambda\ ,\qquad (p_L,p_R)\in\Tilde{\Lambda}\ .
\end{align}
The generator matrix of $\Lambda$ is 
\begin{align}
    M
        =
        \begin{bmatrix}
            \frac{\sqrt{2}}{r}\,\gamma^{-1}& \frac{r}{\sqrt{2}}\,B\\
            0&\frac{r}{\sqrt{2}}\,\gamma^{T}
        \end{bmatrix}.
\end{align}
where $(\gamma)_{i\mu}\equiv e^{\mu}_{i}$.
Each lattice point of $\Lambda$ can be represented as a linear combination of the rows of $M$.

The modular invariance of a Narain CFT guarantees that  the momentum lattice $\Lambda$ is even, i.e., all the lattice vectors have even norm,
\begin{align}
           ^{\forall}\lambda\in\Lambda\ ,\qquad \lambda\circledcirc\lambda\equiv\lambda\,\eta\,\lambda^{T} \in 2\BZ\ ,
\end{align}
and also self-dual, $\Lambda^{*}=\Lambda$, where $\Lambda^{*}$ is the dual lattice of $\Lambda$ defined by
\begin{align}
           \Lambda^{*}=\left\{\lambda'\in \BR^{2n}\ \big|\ ^{\forall}\lambda\in\Lambda,\ \lambda'\circledcirc\lambda\in\BZ\right\}\ .
\end{align}
Conversely, any even self-dual lattice yields the momentum lattice of a modular invariant Narain CFT.

\subsection{Code CFTs}
\label{ss:narain_code}

One can construct Narain CFTs of central charge $n$ from quantum subsystem codes as follows.

First, let us introduce the dual code $\CC^\perp$ of a classical code $\CC$ of length $2n$ with respect to the Lorentzian inner product by
\begin{align}
    \CC^\perp =  \left\{\, c'\in \BF_p^{2n}\,\big|\, c'\,\eta\, c^T = 0~\text{mod}~p\,, ~c \in \CC\,\right\} \ ,
\end{align}
where $\eta$ is defined by
\begin{align}\label{eq:eta-def}
    \eta 
        =
        \left[
        \begin{array}{cc}
    	0 & I_{n} \\
            I_n & 0
        \end{array}
        \right] \ .
\end{align}
The code $\CC$ is called \emph{self-orthogonal} if $\CC\subset\CC^\perp$ and \emph{self-dual} if $\CC = \CC^\perp$.

A classical $[2n,n]_p$ code $\CC$ is Lorentzian self-dual if and only if it is generated by a $n \times 2n$ matrix $G$ satisfying
\begin{align}\label{code_CFT_condition}
    G\,\eta\,G^T = 0\qquad \text{mod}~ p \ .
\end{align}
Note that Lorentzian self-orthogonal $[2n,n]_p$ codes are automatically self-dual (Proposition 3.1 in \cite{Kawabata:2022jxt}).

As discussed around \eqref{eq:CC-stabilizer}, we can associate an $[[n,k,r]]_p$ subsystem code whose gauge group is specified by $\SH_\text{sub}$ with a classical $[2n, n+r-k]_p$ code $\CC_\text{sub}$ over $\BF_p^{2n}$ with generator matrix $G = \SH_\text{sub}$.
Let us consider an $[[n,k,k]]_p$ subsystem code with a gauge group $\SH_\text{sub}$, where $k\le \lfloor n/2\rfloor$. 
If $\SH_\text{sub}$ is subject to the condition
\begin{align}\label{CFT_condition}
     \SH_\text{sub}\,\eta\,\SH_\text{sub}^T = 0\qquad \text{mod}~ p \ ,
\end{align}
then the classical $[2n,n]_p$ code $\CC_\text{sub}$ defined by \eqref{eq:CC-stabilizer} becomes Lorentzian self-dual as the generator matrix $G = \SH_\text{sub}$ of the code $\CC_\text{sub}$ satisfies \eqref{code_CFT_condition}.

Now, we consider the $[2n,n]_p$ Lorentzian self-dual classical code $\CC_\text{sub}$ defined by \eqref{eq:CC-stabilizer} associated with an $[[n,k,k]]_p$ subsystem code with the gauge group $\SH_\text{sub}$ satisfying the condition \eqref{CFT_condition}.
Then, we apply the Construction A \cite{conway2013sphere} to $\CC_\text{sub}$ to define a $2n$-dimensional lattice $\Lambda(\CC_\text{sub})$ :
\begin{align}
   \label{conA}
       \Lambda(\CC_\text{sub})=\left\{\dfrac{c+p\,m}{\sqrt{p}}\ \Bigg|\ c\in\CC_\text{sub}, ~m\in\BZ^{2n} \right\}.
\end{align}
The resulting lattice $\Lambda(\CC_\text{sub})$ is Lorentzian self-dual with the metric $\eta$ since the classical code $\CC_\text{sub}$ is Lorentzian self-dual with respect to the metric $\eta$:
\begin{align}
       \Lambda(\CC_\text{sub}) 
        = \Lambda(\CC_\text{sub})^{*}
        \qquad
        \Leftrightarrow 
        \qquad
        \CC_\text{sub}=\CC_\text{sub}^{\perp} \ .
\end{align}

For odd prime $p$, the Lorentzian self-dual lattice $\Lambda(\CC_\text{sub})$ is also even \cite{Yahagi:2022idq} and thus one can construct a Narain CFT of central charge $n$ from a quantum subsystem code satisfying \eqref{CFT_condition}.

For $p=2$, the Lorentzian self-dual lattice $\Lambda(\CC_\text{sub})$ is even when $\CC_\text{sub}$ is a doubly-even code, i.e., for $c \in\CC_{\mathrm{sub}}$
\begin{align}\label{doubly-even}
    c \circledcirc c\in 4\,\BZ \ .
\end{align}
Equivalently, this condition can be written as $\mathrm{diag}(\SH_{\mathrm{sub}}\,\eta\,\SH_{\mathrm{sub}}^T) = 0$ mod $4$.
Note that for the B-form codes over $\BF_2$ subject to the condition \eqref{B-form_CFT_condition}, $\CC_\text{sub}$, the doubly-evenness is automatically satisfied \cite{Dymarsky:2020qom}.
One can construct a Narain CFT of central charge $n$ from a quantum subsystem code satisfying \eqref{CFT_condition} and \eqref{doubly-even}.

Now we consider the torus partition function of the Narain code CFT constructed from an $[[n,k,k]]_p$ subsystem code.
Using the complete weight enumerator polynomial \eqref{CWE}, the partition function can be written as \cite{Dymarsky:2020qom,Kawabata:2022jxt}
\begin{align}
\label{partition}
    Z_{\CC}(\tau,\Bar{\tau})
        =
        \Tr_\CH\left[q^{L_0-\frac{n}{24}}\,\Bar{q}^{\Bar{L}_0-\frac{n}{24}}\right]
        =
        \dfrac{1}{|\eta(\tau)|^{2n}}\,W_{\CC}(\{\psi^{+}_{ab}\})\ ,
\end{align}
where $\CH$ is the Hilbert space with periodic boundary condition on a circle, and $q\equiv e^{2\pi \i\tau}$ with $\tau$ the modulus of a torus. 
Here, we define the complete weight enumerator polynomial of a classical code $\CC$ as
\begin{align}\label{CWE}
    W_{\CC}(\{x_{ab}\})=\sum_{c\,\in\,\CC}\prod_{(a,b)\,\in\,\BF_p\times\BF_p}x_{ab}^{\text{wt}_{ab}(c)},
\end{align}
where $\text{wt}_{ab}(c)\equiv|\,\{i\in\{1,\cdots,n\}\,|\,(c_i,c_{i+n})=(a,b)\}\,|$ is the composition of each codeword $c\in\CC$.
The Dedekind eta function $\eta(\tau)$ is defined by
\begin{align}
    \eta(\tau)=q^{\frac{1}{24}}\prod_{m=1}^{\infty}(1-q^m)\ ,
\end{align}
and $\psi^{+}_{ab}$ is given by
\begin{align}\label{psi+}
     \psi^{+}_{ab}(\tau,\Bar{\tau})
        =
        \Theta_{a+b,p}(\tau)\,\Bar{\Theta}_{a-b,p}(\Bar{\tau})+\Theta_{a+b-p,p}(\tau)\,\Bar{\Theta}_{a-b-p,p}(\Bar{\tau}) \ ,
\end{align}
where $\Theta_{m,k}(\tau)$ is the theta function:
\begin{align}
    \Theta_{m,k}(\tau)=\sum_{n\in\BZ}\,q^{k\left(n+\frac{m}{2k}\right)^2}\ .
\end{align}

\subsection{$\BZ_2$-gauging of code CFTs}

Let $\CB$ be a bosonic CFT whose momentum lattice is given by the Construction A lattice \eqref{conA} of a Lorentzian self-dual code $\CC_\text{sub}$.
Suppose there exists a lattice vector $\chi \in \Lambda(\CC_\text{sub})$ whose half is not in the lattice, i.e., $\frac{\chi}{2} \not\in \Lambda(\CC_\text{sub})$.
Since $\CC_\text{sub}$ is Lorentzian self-dual, any lattice vector $\lambda \in \CC_\text{sub}$ has an integral inner product with $\chi$, i.e., $\lambda \circledcirc \chi \in \BZ$.
Thus, one can split the lattice vectors in $\Lambda(\CC_\text{sub})$ into two sets as follows:
\begin{align}
    \Lambda(\CC_\text{sub})
        =
        \Lambda_0 \cup \Lambda_1 \ ,
\end{align}
where 
\begin{align}
    \begin{aligned}
        \Lambda_0 
            &=
                \left\{ 
                    \lambda\in \Lambda(\CC_\text{sub}) \, \big|\, \lambda \circledcirc \chi = 0~\text{mod}~2\, \right\} \ , \\
        \Lambda_1
            &=
                \left\{ 
                    \lambda\in \Lambda(\CC_\text{sub}) \, \big|\, \lambda \circledcirc \chi = 1~\text{mod}~2\, \right\} \ .           
    \end{aligned}
\end{align}
Then, this splitting induces the $\BZ_2$ symmetry $\sigma_\chi$ of the (vertex) operators in the CFT $\CB$:
\begin{align}\label{Z2_action_untwisted}
    \sigma_\chi: V_\lambda ~\to~ (-1)^{\lambda\circledcirc\chi}\,V_\lambda \ ,
\end{align}
where $V_\lambda$ is the vertex operator associated with the lattice vector $\lambda\in \Lambda(\CC_\text{sub})$.
From the spin-selection rule in the twisted sector \cite{Lin:2019kpn}, the $\BZ_2$ symmetry $\sigma_\chi$ is shown to be non-anomalous in \cite{Kawabata:2023iss} if the vector $\chi$ satisfies the following condition:
\begin{align}\label{non-anomalous-condition}
    \chi\circledcirc\chi \in 4\,\BZ \ .
\end{align}

It follows from \eqref{conA} that all code CFTs built from Lorentzian self-dual codes through the Construction A have the $\BZ_2$ symmetry associated with the vector $\chi = \sqrt{p}\,{\bf 1}_{2n}$ for any prime $p$.
Furthermore, if a Lorentzian self-dual code contains the all-one vector, the corresponding code CFT has an additional $\widehat\BZ_2$ symmetry associated with the vector $\chi = \frac{1}{\sqrt{p}}\,{\bf 1}_{2n}$.
By taking into account the condition \eqref{non-anomalous-condition}, we find that the $\BZ_2$ symmetry is non-anomalous when $n$ is even while the $\widehat\BZ_2$ symmetry is non-anomalous when $\CC_\text{sub}$ contains ${\bf 1}_{2n}$ as a codeword and $n \in 2p\,\BZ$.

Now, let us place $\CB$ on a circle.
Let $\CH$ and $\CH_{\sigma_\chi}$ be the untwisted and twisted Hilbert spaces whose corresponding operators are subject to the periodic boundary condition and twisted one by the $\BZ_2$ symmetry along the circle, respectively.
We then introduce the $\BZ_2$-graded partition functions on a torus as follows:
\begin{align}
    \begin{aligned}
        S 
            &\equiv
            \Tr_\CH\left[\frac{1+\sigma_\chi}{2}\,q^{L_0-\frac{n}{24}}\,\Bar{q}^{\Bar{L}_0-\frac{n}{24}}\right] \ , \\
        T
            &\equiv
            \Tr_\CH\left[\frac{1-\sigma_\chi}{2}\,q^{L_0-\frac{n}{24}}\,\Bar{q}^{\Bar{L}_0-\frac{n}{24}}\right] \ , \\
        U 
            &\equiv
            \Tr_{\CH_{\sigma_\chi}}\left[\frac{1+\sigma_\chi}{2}\,q^{L_0-\frac{n}{24}}\,\Bar{q}^{\Bar{L}_0-\frac{n}{24}}\right] \ , \\
        V
            &\equiv
            \Tr_{\CH_{\sigma_\chi}}\left[\frac{1-\sigma_\chi}{2}\,q^{L_0-\frac{n}{24}}\,\Bar{q}^{\Bar{L}_0-\frac{n}{24}}\right] \ .
    \end{aligned}
\end{align}
Compared with \eqref{partition}, the partition function of the bosonic code CFT $\CB$ constructed from the code $\CC$ is written as
\begin{align}
    Z_\CC = S + T \ .
\end{align}

When the $\BZ_2$ symmetry $\sigma_\chi$ is non-anomalous, the orbifolded theory $\CO$ can be constructed from $\CB$ by gauging the symmetry.
The resulting theory no longer possesses $\sigma_\chi$, but has a new $\BZ_2$ symmetry $\hat\sigma_\chi$, which allows us to introduce the $\BZ_2$-grading of the operators as well as the untwisted and twisted sectors.
Each sector in the orbifolded theory $\CO$ is in correspondence with one of the four sectors in the original theory $\CB$ as shown 
in Table \ref{table:gauging}.
Thus, the partition function $Z_\CC^\CO$ of the untwisted sector in the orbifolded theory is given by
\begin{align}
    Z_\CC^\CO = S + U \ .
\end{align}
Similarly, one can construct a fermionic CFT $\CF$ from $\CB$ by fermionization, i.e., coupling $\CB$ with the Kitaev Majorana chain and gauging the diagonal $\BZ_2$ subgroup of $\sigma_\chi$ and the $\BZ_2$ symmetry of the Majorana chain~\cite{Kitaev:2000nmw}.
The fermionic operators are subject to either periodic (P) or anti-periodic (A) boundary condition along each cycle of the torus.
Thus, there are four choices of a spin structure on the torus:
\begin{align}
    \mathrm{NS}: (\mathrm{A},\mathrm{A}),\quad \widetilde{\mathrm{NS}}: (\mathrm{A},\mathrm{P}),\quad
    \mathrm{R}: (\mathrm{P},\mathrm{A}),\quad \widetilde{\mathrm{R}}: (\mathrm{P},\mathrm{P}),
\end{align}
where $(x,y)$ stands for the boundary conditions $x$ and $y$ along the spacial and timelike cycles, respectively.

Using the fermionic parity $(-1)^F$, the partition functions in the four sectors are given by
\begin{align}
    \begin{aligned}
            Z_\CC^\text{NS} 
                &\equiv
                    \Tr_\text{NS}\left[ q^{L_0-\frac{n}{24}}\,\Bar{q}^{\Bar{L}_0-\frac{n}{24}}\right] &
                &=
                    S + V \ , \\
            Z_\CC^{\widetilde{\text{NS}} }
                &\equiv
                    \Tr_\text{NS}\left[ (-1)^F\, q^{L_0-\frac{n}{24}}\,\Bar{q}^{\Bar{L}_0-\frac{n}{24}}\right] &
                &=
                    S - V \ , \\
            Z_\CC^\text{R} 
                &\equiv 
                    \Tr_\text{R}\left[ q^{L_0-\frac{n}{24}}\,\Bar{q}^{\Bar{L}_0-\frac{n}{24}}\right] &
                &=
                    T + U \ , \\
            Z_\CC^{\widetilde{\text{R}}}
                &\equiv 
                    \Tr_\text{R}\left[ (-1)^F\, q^{L_0-\frac{n}{24}}\,\Bar{q}^{\Bar{L}_0-\frac{n}{24}}\right] &
                &=
                    T - U \ .
    \end{aligned}
\end{align}
The correspondences between the sectors in the bosonic theory $\CB$ and those in the fermionized theory $\CF$ are summarized in Table \ref{table:gauging}.

\begin{table}
  \centering
      \begin{subtable}[t]{0.3\textwidth}
        \centering
              \begin{tabular}{ccc}
              \toprule
                $\CB$  & untwisted  &  twisted  \\
                \hline 
                even  & $S$  & $U$ \\
                odd  & $T$   & $V$ \\\bottomrule
              \end{tabular}
        \caption{Bosonic code CFT}
      \end{subtable}
      ~~~~
      \begin{subtable}[t]{0.3\textwidth}
            \centering
                  \begin{tabular}{ccc}
                  \toprule
                    $\CO$  & untwisted  &  twisted  \\
                    \hline 
                    even  & $S$  & $T$ \\
                    odd  & $U$   & $V$ \\\bottomrule
                  \end{tabular}
            \caption{Orbifolded code CFT}
      \end{subtable}
      ~~~~
        \begin{subtable}[t]{0.3\textwidth}
            \centering
              \begin{tabular}{ccc}
              \toprule
                $\CF$  & NS sector  &  R sector  \\
                \hline 
                even  & $S$  & $T$ \\
                odd  & $V$   & $U$ \\\bottomrule
              \end{tabular}
            \caption{Fermionized code CFT}
        \end{subtable}
  \vspace{0.5cm}
  \caption{The partition functions of the $\BZ_2$ graded sectors in the bosonic code theory $\CB$, the orbifolded theory $\CO$, and the fermionized theory $\CF$. The gradings are given by the $\BZ_2$ symmetry $\sigma$ for $\CB$, its dual symmetry $\hat{\sigma}$ for $\CO$, and the fermion parity $(-1)^F$ for $\CF$, respectively.
  The other fermionized theory $\widetilde{\CF}$ can be constructed by stacking $\CF$ with the Kitaev Majorana chain, which amounts to swapping the even $(T)$ and odd $(U)$ sectors in the R sector of $\CF$.}
  \label{table:gauging}
\end{table}

For code CFTs, the partition functions in the four sectors can be written by using the complete weight enumerator polynomial as follows \cite{Kawabata:2023iss}:
\begin{align}
    S
        &=
        \dfrac{1}{2\,|\eta(\tau)|^{2n}}\,\left[ W_{\CC}(\{\psi^{+}_{ab}\}) + W_{\CC}(\{\psi^{-}_{ab}\})\right]\ , \\
    T
        &=
        \dfrac{1}{2\,|\eta(\tau)|^{2n}}\,\left[ W_{\CC}(\{\psi^{+}_{ab}\}) - W_{\CC}(\{\psi^{-}_{ab}\})\right]\ , \\
    U
        &=
        \dfrac{1}{2\,|\eta(\tau)|^{2n}}\,\left[ W_{\CC}(\{\tilde\psi^{+}_{ab}\}) + W_{\CC}(\{\tilde\psi^{-}_{ab}\})\right]\ , \\
    V
        &=
        \dfrac{1}{2\,|\eta(\tau)|^{2n}}\,\left[ W_{\CC}(\{\tilde\psi^{+}_{ab}\}) - W_{\CC}(\{\tilde\psi^{-}_{ab}\})\right]\ ,
\end{align}
where $\psi^+_{ab}$ is defined by \eqref{psi+} while the functions $\psi^-_{ab}, \, \tilde \psi^+_{ab},\,  \tilde\psi^-_{ab}$ depend on the choice of $\chi$ as shown in the following paragraphs.

\paragraph{$\BZ_2$ gauging for non-$\BF_4$-even codes}

Non-$\BF_4$-even codes do not contain the all-one vector ${\bf 1}_{2n}$, but their code CFTs have the $\BZ_2$ symmetry associated with the vector:
\begin{align}\label{nonF4even_Z2}
    \chi = \sqrt{2}\,{\bf 1}_{2n} \ .
\end{align}
The $\BZ_2$ symmetry is non-anomalous for any $n$.

\begin{align}
    \begin{aligned}
        \psi^{-}_{ab}(\tau,\Bar{\tau})
            &=
            (-1)^{a+b}\left[ \Theta_{a+b,2}(\tau)\,\Bar{\Theta}_{a-b,2}(\Bar{\tau}) + \Theta_{a+b-2,2}(\tau)\,\Bar{\Theta}_{a-b-2,2}(\Bar{\tau}) \right] \ , \\
        \tilde\psi^{+}_{ab}(\tau,\Bar{\tau})
            &=
             \Theta_{a+b,2}(\tau)\,\Bar{\Theta}_{a-b-2,2}(\Bar{\tau}) + \Theta_{a+b-2,2}(\tau)\,\Bar{\Theta}_{a-b,2}(\Bar{\tau}) \ , \\  
        \tilde\psi^{-}_{ab}(\tau,\Bar{\tau})
            &=
             (-1)^{(a+1)(b+1)}\left[ \Theta_{a+b,2}(\tau)\,\Bar{\Theta}_{a-b-2,2}(\Bar{\tau}) - \Theta_{a+b-2,2}(\tau)\,\Bar{\Theta}_{a-b,2}(\Bar{\tau}) \right]\ .
    \end{aligned}
\end{align}

\paragraph{$\BZ_2$ gauging for $\BF_4$-even codes}
$\BF_4$-even codes contain the all-one vector ${\bf 1}_{2n}$, and their code CFTs have the $\widehat\BZ_2$ symmetry associated with the vector:
\begin{align}\label{F4even_Z2}
    \chi = \frac{1}{\sqrt{2}}\,{\bf 1}_{2n} \ .
\end{align}
The $\widehat\BZ_2$ symmetry is non-anomalous only when $n\in 4\BZ$.

\begin{align}
    \begin{aligned}
        \psi^{-}_{ab}(\tau,\Bar{\tau})
            &=
            (-1)^{\frac{a+b}{2}}\left[ \Theta_{a+b,2}(\tau)\,\Bar{\Theta}_{a-b,2}(\Bar{\tau}) - \Theta_{a+b-2,2}(\tau)\,\Bar{\Theta}_{a-b-2,2}(\Bar{\tau}) \right] \ , \\
        \tilde\psi^{+}_{ab}(\tau,\Bar{\tau})
            &=
             \Theta_{a+b+1,2}(\tau)\,\Bar{\Theta}_{a-b,2}(\Bar{\tau}) + \Theta_{a+b-1,2}(\tau)\,\Bar{\Theta}_{a-b-2,2}(\Bar{\tau}) \ , \\  
        \tilde\psi^{-}_{ab}(\tau,\Bar{\tau})
            &=
             e^{\pi\i\,\left( a + \frac{1}{2}\right) \left( b + \frac{1}{2}\right)}\left[ \Theta_{a+b+1,2}(\tau)\,\Bar{\Theta}_{a-b,2}(\Bar{\tau}) - \Theta_{a+b-1,2}(\tau)\,\Bar{\Theta}_{a-b-2,2}(\Bar{\tau}) \right]\ .
    \end{aligned}
\end{align}

\paragraph{$\BZ_2$ gauging for general $p$-ary codes}
The code CFTs have the $\BZ_2$ symmetry associated with the vector:
\begin{align}\label{nonF4like_Z2}
    \chi = \sqrt{p}\,{\bf 1}_{2n} \ .
\end{align}
The $\BZ_2$ symmetry is non-anomalous only when $n\in 2\BZ$.

\begin{align}
    \begin{aligned}
        \psi^{-}_{ab}(\tau,\Bar{\tau})
            &=
            (-1)^{a+b}\left[ \Theta_{a+b,p}(\tau)\,\Bar{\Theta}_{a-b,p}(\Bar{\tau}) - \Theta_{a+b-p,p}(\tau)\,\Bar{\Theta}_{a-b-p,p}(\Bar{\tau}) \right] \ , \\
        \tilde\psi^{+}_{ab}(\tau,\Bar{\tau})
            &=
             \Theta_{a+b,p}(\tau)\,\Bar{\Theta}_{a-b-p,p}(\Bar{\tau}) + \Theta_{a+b-p,p}(\tau)\,\Bar{\Theta}_{a-b,p}(\Bar{\tau}) \ , \\  
        \tilde\psi^{-}_{ab}(\tau,\Bar{\tau})
            &=
             e^{\frac{2\pi\i}{p}\left( a + \frac{p}{2}\right)\left( b + \frac{p}{2}\right)}\left[ \Theta_{a+b,p}(\tau)\,\Bar{\Theta}_{a-b-p,p}(\Bar{\tau}) - \Theta_{a+b-p,p}(\tau)\,\Bar{\Theta}_{a-b,p}(\Bar{\tau}) \right]\ .
    \end{aligned}
\end{align}

When $p$-ary codes contain ${\bf 1}_{2n}$, their code CFTs have the $\widehat\BZ_2$ symmetry associated with the vector:
\begin{align}\label{F4like_Z2}
    \chi = \frac{1}{\sqrt{p}}\,{\bf 1}_{2n} \ .
\end{align}
The $\widehat\BZ_2$ symmetry is non-anomalous only when $n\in 2p\,\BZ$.
It follows from \eqref{Z2_action_untwisted} that the action of the $\widehat\BZ_2$ symmetry on the untwisted sector is the same as that of the $\BZ_2$ symmetry associated with the vector \eqref{nonF4like_Z2} when $p$ is odd.
In general, when a $\BZ_2$-grading is given on the untwisted sector, the corresponding twisted sector is determined without ambiguity \cite{Vafa:1986wx,Vafa:1994rv}.
Thus, the gauging by the $\BZ_2$ and $\widehat\BZ_2$ symmetries for odd prime $p$ yields the same orbifolded and fermionized theories.

\section{Construction of $\CN=1,2$ supercurrents}
\label{ss:fermion}

In this section, we discuss supersymmetry in fermionic code CFTs (i.e., the fermionized theories of code CFTs).
To show the existence of supersymmetry, we have to construct the supercurrent $G(z)$, which is a generator of superconformal transformation.
The supercurrent $G(z)$ is a Virasoro primary operator characterized by its conformal dimension $(h,\bar{h}) = (3/2,0)$ and its operator product expansion (OPE)
\begin{align}
\label{eq:supcr_def}
    G(z) \, G(w) \sim \frac{2n/3}{(z-w)^3} + \frac{2 \,T(w)}{z-w}\,.
\end{align}
Generally, its construction is difficult due to a relative phase in correlation functions of vertex operators (see~\cite{Harvey:2020jvu,Moore:2023zmv} for a recent discussion in some models).
However, there are general criteria for a fermionic CFT to have supersymmetry, called supersymmetry conditions~\cite{Bae:2021jkc,Bae:2021lvk}.
\begin{itemize}
    \item The NS sector contains spin-$3/2$ primary operators.
    \item The R sector satisfies the positive energy condition $E_R\geq0$.
    \item The torus partition function with spin structure $\widetilde{\mathrm{R}}$ is constant.
\end{itemize}
The first condition ensures the presence of primaries with the same spin as for the supercurrent.
The second corresponds to the supersymmetric unitarity condition and the third imposes that the $\widetilde{\mathrm{R}}$ partition function is consistent with the Witten index.
Although these criteria are necessary but not sufficient for supersymmetry, they strongly support the existence of supersymmetry.
This condition is useful for detecting whether a given theory is supersymmetric since all conditions can be easily computed for code CFTs~\cite{Kawabata:2023usr}.

For certain fermionic code CFTs, we can show the existence of supersymmetry.
To show it, we have to construct a supercurrent from operator contents in the theory.
A fermionic code CFT is defined by a set of vertex operators based on an odd self-dual lattice $\Lambda_{\mathrm{NS}}$ and its holomorphic stress tensor is given by
\begin{align}
\label{eq:stress}
    T(z) = -\frac{1}{2}\sum_{i=1}^n :\partial X_L^i(z)\,\partial X_L^i(z):\,.
\end{align}
If there exists a supercurrent, it should be constructed from a linear combination of holomorphic vertex operators $V_{K_l}(z)= :e^{\i K_l\cdot X_L(z)}:$ with conformal dimension $(h,\bar{h}) = (3/2,0)$:
\begin{align}
\label{eq:cand_sp}
    G(z) = \sum_{i}\, a_l \,V_{K_l}(z)\,,
\end{align}
where $K_l$ is the left-moving momenta such that $K_l\cdot K_l=3$. 
Note that their right-moving momenta are vanishing since we are considering operators with $\bar{h}=0$.

Our statement is the following: 
Let a fermionic CFT with central charge $n$ be constructed from an odd self-dual lattice $\Lambda_{\mathrm{NS}}\subset \BR^{2n}$. Suppose that the theory does not have any vertex operator with conformal dimension $(h,\bar{h})= (1,0),(2,0)$, while it contains $2n$ vertex operators with $(h,\bar{h})= (3/2,0)$.
(Essentially, it is only necessary that the left-moving momenta of vertex operators with $(h,\bar{h})= (3/2,0)$ are orthogonal.)
Then, a supercurrent satisfying \eqref{eq:supcr_def} can be constructed from the set of vertex operators with $(h,\bar{h})= (3/2,0)$. Furthermore, we can construct $\CN=2$ supercurrents, which show the existence of $\CN=2$ supersymmetry in the left-moving sector.
If the theory only contains $2N(<2n)$ vertex operators with $(h,\bar{h})= (3/2,0)$, the theory has a supersymmetric subsector with smaller central charge $N$.
For the right-moving sector, the same follows.

This is expected since the supercurrent OPE \eqref{eq:supcr_def} requires that no operators appear except for the identity and the stress tensor and we assume that there are no vertex operators with $(h,\bar{h})= (1,0),(2,0)$, which might appear in the OPE.
Note that this statement ensures the existence of supersymmetry, although the supersymmetry condition only imposes necessary conditions for supersymmetry.

Let us see that the statement holds.
To consider the OPE of $G(z)$, we start with the OPE of the vertex operators
\begin{align}
\begin{aligned}
    :V_{K_l}(z):\;:V_{K_m}(w): \;=\; (z-w)^{K_l\cdot K_m} :V_{K_l}(z)V_{K_m}(w):\,,
\end{aligned}
\end{align}
where 
\begin{align}
    \begin{aligned}
        :V_{K_l}(z)V_{K_m}(w): \;&=\;:\,e^{i(K_l+K_m)\cdot X_L(w)}:
    +\i (z-w) : K_l \cdot \partial X_L(w) \,e^{i(K_l+K_m)\cdot X_L(w)}:\\
    &+\frac{1}{2}(z-w)^2 :\left(\i K_l \cdot \partial^2 X_L(w) - (K_l \cdot \partial X_L(w))^2\right) \,e^{i(K_l + K_m)\cdot X_L(w)}:
    \end{aligned}
\end{align}
There are some possibilities for the value of $K_l\cdot K_m$.
From the spin-statics theorem, the holomorphic vertex operators $V_{K_l}(z)$ with spin $s=3/2$ are fermionic. After taking their OPEs, we obtain only bosonic holomorphic vertex operators with an integral conformal weight $h\in\BZ_{\geq0}$. Therefore, the OPE does not have a branch cut, and the inner product $K_l\cdot K_m$ can take the following values:
\begin{align}
    K_l\cdot K_m =
    \begin{dcases}
        -3 & \text{when } K_l = -K_m\\
        -2 & \text{when } (K_l+K_m)^2=2\\
        -1 & \text{when } (K_l+K_m)^2=4 
    \end{dcases}
\end{align}
where we only consider negative values of $K_l\cdot K_m$ since such terms appear as a singular term in the OPE.
The last two cases occur only when there exists a holomorphic vertex operator $V_{K_l+K_m}(z)$ with conformal weight $h=(1,0),(2,0)$.
However, we assume that there are no such operators.
Thus, the first case is only possible and the OPE becomes
\begin{align}
    :V_{K_l}(z):\;:V_{-K_l}(w): \;\sim\; \frac{1}{(z-w)^3} +\frac{\i K_l\cdot \partial X_L(w)}{(z-w)^2} + \frac{1/2}{z-w}\left(\i K_l \cdot \partial^2 X_L(w)\, - :(K_l \cdot \partial X_L(w))^2:\right) 
\end{align}
Note that if we have a vertex operator with $(p_L,p_R) = (K_l,0)$, then we also have $(p_L,p_R) = (-K_l,0)$ since the eigenvalues form a lattice.
In total, we have $2N$ holomorphic vertex operators with $(h,\bar{h}) = (3/2,0)$. 
They are in pair as $K_1=-K_{N+1},K_2=-K_{N+2}, \cdots,K_N=-K_{2N}$.
We can easily show that $K_l$ $(l=1,2,\cdots,N)$ are orthogonal because there are no holomorphic vertex operators with weight $h=(1,0),(2,0)$. By using this, the candidate supercurrent \eqref{eq:cand_sp} has the OPE
\begin{align}
\begin{aligned}
    G(z) \,G(w) &= \sum_{l=1}^N a_l\, a_{N+l}\, \left(V_{K_l}(z) V_{K_{N+l}} (w) + V_{K_{N+l}}(z)V_{K_l}(w)\right) \,,\\
    &\sim \frac{2}{(z-w)^3} \sum_{l=1}^N a_l\,a_{N+l}
    -\frac{1}{z-w} \sum_{l=1}^N a_l\,a_{N+l} \,K_l^i \,K_l^j\,
    :\partial X_L^i(w)\partial X_L^j(w):\,.
\end{aligned}
\end{align}
Since the stress tensor in the theory is given by~\eqref{eq:stress},
to match the above OPE with~\eqref{eq:supcr_def}, we obtain the two constraints
\begin{align}
\sum_{l=1}^N a_l \,a_{N+l} = \frac{n}{3}\,,\qquad 
    \sum_{l=1}^N a_l\,a_{N+l} \,K_l^i \,K_l^j = \delta^{ij}\,,
\end{align}
where we are considering the theory with central charge $n$.
To solve the two constraints, we assume $a_1 a_{N+1} = a_2 a_{N+2} = \cdots = a_N a_{2N} $.  

When $N=n$, the first constraint gives $a_l a_{N+l} = 1/3$. The second condition reduces to
\begin{align}
    \sum_{l=1}^n \frac{K_l^i}{\sqrt{3}}\,\frac{K_l^j}{\sqrt{3}} = \delta^{ij}\,,
\end{align}
where each $K_l$ is orthogonal in the Euclidean space $\BR^n$. By using an orthogonal transformation $P\in \mathrm{O}(n)$, we can replace $K_l/\sqrt{3}$ into a vector $e_l$ of the standard basis in $\BR^n$.
Then, it is obvious that the standard basis satisfies $\sum_{l=1}^n e_l^i \,e_l^j = \delta^{ij}$ and the constraints are solved.
This implies that when $N=n$, the supercurrent is given by
\begin{align}
\label{eq:supercurrent}
    G(z) = \frac{1}{\sqrt{3}}\sum_{l=1}^{2n} V_{K_l}(z)\,.
\end{align}
Furthermore, we can construct the $\CN=2$ superconformal algebra.
Let us set the generators
\begin{align}
\begin{aligned}
    T(z) = -\frac{1}{2}\,\sum_{i=1}^{n-1} :\partial X_L^i(z)\,\partial X_L^i(z):\,,\quad
    J(z) = \frac{\i}{3} \sum_{l=1}^n K_l\cdot \partial X_L(z)\,,\\
    G^+(z) = \sqrt{\frac{2}{3}}\,\sum_{l=1}^n V_{K_l}(z) \,,\qquad
    G^-(z) = \sqrt{\frac{2}{3}}\,\sum_{l=1}^n V_{-K_l}(z)\,.
    \end{aligned}
\end{align}
Then, these yield the $\CN=2$ superconformal algebra. For example, we have
\begin{align}
\label{eq:N=2_susy}
\begin{aligned}
    G^+(z)\,G^-(w) \sim \frac{2n/3}{(z-w)^3} + \frac{2J(w)}{(z-w)^2} + \frac{2T(w) + \partial J(w)}{(z-w)}\,,\\
    J(z) \,G^\pm(w) = G^\pm(w) \,,\qquad J(z)\,J(w) = \frac{n/3}{(z-w)^2}\,.
\end{aligned}
\end{align}
This can be understood as the generalization of the free boson construction of the $\CN=2$ superconformal unitary minimal model with $k=1$~\cite{Waterson:1986ru} to larger central charges.

When $N<n$, the theory does not have the superconformal OPE for the stress tensor~\eqref{eq:stress}, but there is a supersymmetric subsector with a smaller central charge $N<n$. 
For example, if all $K_l$s ($l=1,2,\cdots,2n-2)$ are orthogonal to the $n$-th direction of $\BR^n$. Then, the superconformal algebra with central charge $n-1$ is generated by the supercurrent
\begin{align}
    G(z) = \frac{1}{\sqrt{3}}\sum_{l=1}^{2n-2} V_{K_l}(z)\,,
\end{align}
and the stress tensor is
\begin{align}
    T(z) = -\frac{1}{2}\,\sum_{i=1}^{n-1} :\partial X_L^i(z)\,\partial X_L^i(z):\,.
\end{align}
Note that this stress tensor is not \eqref{eq:stress} with central charge $n$.

\paragraph{Example $(p=3)$.}
We consider an example with supersymmetry from a ternary subsystem code generated by
\begin{align}
    \SH_{\mathrm{sub}} =
    \left[
    \begin{array}{cc|cc}
        ~1~ & ~0~ & ~0~ & ~-1~\\
        ~0~ & ~1~ & ~1~ & ~0~
    \end{array}\right]\,.
\end{align}
The weight enumerator of this code is $W_\CC({x_{ab}}) = x_{00}^2 + x_{01} x_{10} + x_{01} x_{20} + x_{02} x_{10} + x_{02} x_{20} + x_{11} x_{12} + x_{11} x_{21} + x_{12} x_{22} + x_{21} x_{22}$.
We can easily check that the corresponding CFT fermionized using~\eqref{nonF4like_Z2} satisfies all the supersymmetry conditions.
For the first condition, we see the NS lattice theta function
\begin{align}
\label{eq:nslat_exm}
    \Theta_{\mathrm{NS}}(\tau,\bar{\tau}) \supset 1 + 4q^{3/2} + 4\bar{q}^{3/2} + 4q^{3} + 4\bar{q}^{3} +\dots\,,
\end{align}
where we pick up the holomorphic and anti-holomorphic terms. Since it contains $4$ vertex operators with spin-$3/2$, the theory satisfies the first condition.
Additionally, the second condition is satisfied:
\begin{align}
    Z^{\mathrm{R}}_\CC(\tau,\bar{\tau}) = 4 + 16\,(q\bar q)^\frac{1}{3} + 16\,q + 16\,\bar q
                    + 16\,(q\bar q)^\frac{2}{3} 
                    + \cdots\ .
\end{align}
Finally, the $\widetilde{\mathrm{R}}$ partition function becomes
\begin{align}
    Z^{\widetilde{\mathrm{R}}}_\CC(\tau,\bar{\tau}) = -4 \ .
\end{align}
These indicate the existence of supersymmetry for the fermionic theory.

In this case, we can construct its supercurrent explicitly and prove the existence of supersymmetry.
We should note that there are no holomorphic vertex operators with conformal weight $(h,\bar{h})=(1,0),(0,1)$, $(2,0),(0,2)$ in \eqref{eq:nslat_exm}.
Also, there are $4$ vertex operators with $(h,\bar{h})=(3/2,0)$.
More explicitly, they are the vertex operators $V_{K_l}(z) = :e^{\i K_l \cdot X_L}:$ where
\begin{align}
    K_1 = \sqrt{\frac{3}{2}}\,(1,1),\quad
    K_2 = \sqrt{\frac{3}{2}}\,(1,-1)\,,
\end{align}
and $K_3=-K_1$, $K_4=-K_2$.
Thus, the $\CN=1$ supercurrent is given by \eqref{eq:supercurrent}.

Also, we can construct the $\CN=2$ superconformal algebra by
\begin{align}
    G^+(z) = \sqrt{\frac{2}{3}}\,\left(V_{K_1}(z) + V_{K_2}(z)\right)\,,\quad  G^-(z) = \sqrt{\frac{2}{3}}\,\left(V_{-K_1}(z) + V_{-K_2}(z)\right)\,, 
\end{align}
where the R-symmetry current is
\begin{align}
    J(z) = \i \sqrt{\frac{2}{3}}\;\partial X^1_L(z)\,.
\end{align}
Correspondingly, the NS partition function can be decomposed into the characters of the $\CN=2$ unitary minimal model with $k=1$:
\begin{align}
    Z^{\mathrm{NS}}_\CC(q,\bar{q}) = \left( (\chi_0^{0,0} +\chi_0^{0,2})(\bar\chi_0^{0,0} +\bar\chi_0^{0,2})+2(\chi_1^{1,0}+\chi_1^{1,2})(\bar\chi_1^{1,0}+\bar\chi_1^{1,2})\right)^2\,,
\end{align}
where the characters are given by
\begin{align}
\begin{aligned}
    \chi_0^{0,0}(\tau) = \frac{\Theta_{0,6}(\tau)}{\eta(\tau)}\,,\quad 
    \chi_0^{0,2}(\tau) = \frac{\Theta_{6,6}(\tau)}{\eta(\tau)}\,,\\
    \chi_1^{1,0}(\tau) = \frac{\Theta_{2,6}(\tau)}{\eta(\tau)}\,,\quad 
    \chi_1^{1,2}(\tau) = \frac{\Theta_{4,6}(\tau)}{\eta(\tau)}\,.
\end{aligned}
\end{align}
The fermionized theory is two copies of the $\CN=2$ supersymmetric unitary minimal model with $k=1$ whose partition function is a modular covariant combination of their characters.

\section{Enumeration of B-form codes and their code CFTs}
\label{ss:enumeration}
In this section, we will focus on an $[[n,k,k]]_p$ subsystem code characterized by an $n\times 2n$ matrix $\SH_\text{sub}$ of the form:
\begin{align}\label{B-form_code}
    \SH_\text{sub}
        =
        \left[
            \begin{array}{c|c}
        	~I_n ~& ~B~ 
            \end{array}
        \right] \ ,
\end{align}
where $B$ is an $n\times n$ matrix over $\BF_p$.\footnote{The matrix $\SH_\text{sub}$ of the form \eqref{B-form_code} is not necessarily of the standard form \eqref{subsystem_H}, hence it does not necessarily satisfy the condition \eqref{subsystem_condition}.
On the other hand, it is always possible to make $\SH_\text{sub}$ be of the standard form by changing the basis of the generators of the gauge group \cite{haah2013lattice}.
In the standard form, the parameters $n, k$ for the $[[n,k,k]]_p$ subsystem code associated with \eqref{B-form_code} can be read off from the number of stabilizer generators and the other gauge generators.
}
This type of quantum code is known as the B-form codes \cite{Dymarsky:2020qom}.
The B-form codes satisfy the condition \eqref{CFT_condition} if the matrix $B$ is anti-symmetric:
\begin{align}\label{B-form_CFT_condition}
    B = - B^T \quad \text{mod}~ p \ .
\end{align}
There are infinitely many B-form codes whose corresponding classical codes have the generator matrices of the form \eqref{B-form_code} subject to the condition \eqref{B-form_CFT_condition}.
For example, the Pless symmetry codes and quadratic double circulant codes can be employed as the classical codes corresponding to the B-form codes. 
See appendix \ref{ap:Pless_QDC} for their definitions and details.

We can classify the anti-symmetric B-form codes by using the adjacency matrix of an oriented weighted graph. An oriented graph $G=(V,E)$ is a set of vertices denoted by $V$ and oriented edges denoted by $E$. If $(i,j)\in E\ i,j\in V$, then there exists an edge oriented from vertex $i$ to $j$. A weighted graph is a graph whose edges have weight.
Given an oriented weighted graph $G=\{V,E\}$, the adjacency matrix $B$ is uniquely defined as follows:
\begin{align}
    B_{ij}=\begin{cases}
        W_{ij} & \text{if}\ \  (i,j)\in E\\
        -W_{ij}& \text{if}\ \  (j,i)\in E\\
        0& \text{otherwise}
    \end{cases}
\end{align}
where $W_{ij}$ is weight of the edge $(i,j)\in E$. We assign each weight to an integer in $\{-(p-1)/2,\cdots,0,\cdots,(p-1)/2\}$ for an odd prime $p$. Therefore, the adjacency matrix $B$ of an oriented weighted graph is anti-symmetric and give an anti-symmetric B-form code.
For $p=2$, anti-symmetric matrices are equivalent to symmetric matrices mod $2$, and our classification of the B-form codes reduces to the one using unoriented graphs as in \cite{Dymarsky:2020qom}.

\subsection{$n=1$}

We start with the simplest graph with one node:
\tikzstyle{node}=[draw=black,circle,inner sep=0,minimum size=10]
\begin{center}
    \begin{tikzpicture}[scale=4]
    \node (v1) at (0,0) [node]{1} ;
    \end{tikzpicture}
\end{center}
The generator matrix of the code $B^{(1,p)}$ associated with this graph is given by
   \begin{align}
       \SH_{B^{(1,p)}}=\begin{bmatrix}
           ~1&\Big|&0~
       \end{bmatrix} \ .
   \end{align}
Hence, this is a subsystem code $[[1,0,0]]_p$ associated with a gauge group $\CG=\langle\omega I, X \rangle$. 
Since the gauge group is abelian, this code is actually a stabilizer code $[[1,0]]_p$.

We can easily check this matrix satisfies the condition (\ref{CFT_condition}) to give a modular-invariant CFT. 
The $[2,1]_p$ classical code $\CC_{B^{(1,p)}}$ generated by $\SH_{B^{(1,p)}}$ is 
\begin{align}
    \CC_{B^{(1,p)}} 
        =
            \{(0,0),\, (1,0),\, \cdots,\, (p-1,0)\}\ .
\end{align}
Its complete weight enumerator is
\begin{align}
    W_{B^{(1,p)}}(\{x_{ab}\})
        =
            \sum_{i=0}^{p-1}\, x_{i0}\ .
\end{align}
Thus, the partition function of the code CFT constructed from the code $B^{(1,p)}$ is 
\begin{align}\label{n=1PF}
    Z_{B^{(1,p)}}(\tau, \bar\tau)
        =
            \dfrac{1}{|\eta(\tau)|^2}\,\sum_{i=0}^{p-1}\left[ \Theta_{i}\Bar{\Theta}_{i}+\Theta_{i+p}\Bar{\Theta}_{i+p}\right] \ ,
\end{align}
where we use short-hand notation for the theta functions, $\Theta_{i} \equiv \Theta_{i,p}(\tau)$, $\bar\Theta_{i} \equiv \bar\Theta_{i,p}(\bar\tau)$.

We can see that the resulting code CFT is a free compact boson of radius $R = \sqrt{2p}$ (or $\sqrt{2/p}$ by the T-duality).
To see this, the Construction A lattice takes the form
\begin{align}
    \Lambda\left(\CC_{B^{(1,p)}}\right)
        =
        \left\{ \left(  \frac{m_1}{\sqrt{p}},\, \sqrt{p}\,m_2 \right) \, \bigg|\, m_1, m_2 \in \BZ\right\} \ .
\end{align}
Then, the left- and right-momenta can be read off as
\begin{align}
        p_L = \sqrt{\frac{p}{2}}\,m_2 + \frac{m_1}{\sqrt{2p}} \ , \qquad 
        p_R =  \sqrt{\frac{p}{2}}\,m_2 - \frac{m_1}{\sqrt{2p}} \ ,
\end{align}
which is the momentum lattice of a free compact boson of radius $R = \sqrt{2/p}$.

We note that the case with $p=2$ has been given in \cite{Dymarsky:2020qom}.
Code CFTs with $n=1$ have also been studied in \cite{Alam:2023qac}, where a single compact boson theory whose radius square is any rational number is constructed from quantum stabilizer codes over a finite ring.
For $p=2$, the B-form code is non-$\BF_4$-even and the $\BZ_2$ symmetry generated by $\chi$ can be gauged, which has already considered in~\cite{Dymarsky:2020qom,Kawabata:2023iss}.
For $p=3$, we find that the $B^{(1,3)}$ code CFT is equivalent to the $\BZ_4$ parafermion theory described by the $\SU (2)_4^{D_4}/\U (1)$ coset theory with $D_4 \times \BZ_2$ global symmetry that has a mixed anomaly (see also \cite{Thorngren:2021yso}).

\subsection{$n=2$}

\subsubsection{$p=2$}

For $p=2$, there are two unoriented graphs as shown in Table \ref{tab:n2p2}.

\tikzstyle{node}=[draw=black,circle,inner sep=0,minimum size=10]
\begin{table}[t]
    \centering
    \renewcommand{\arraystretch}{1.5}
    \begin{tabular*}{10cm}{c@{\extracolsep{\fill}}cc}
        \toprule
       Name     & $B_{0}^{(2,2)}$ & $B_{1}^{(2,2)}$ \\\midrule \\[-0.4cm]
       Graph    &   \begin{tikzpicture}[scale=1.5, thick]
                        \node (v1) at (0,0) [node]{} ;
                        \node (v2) at (1,0) [node]{} ;
                    \end{tikzpicture}
                &   \begin{tikzpicture}[scale=1.5, thick]
                        \node (v1) at (0,0) [node]{} ;
                        \node (v2) at (1,0) [node]{} ;
                        \draw[-](v1)--(v2);
                    \end{tikzpicture} \\  [10pt] 
        $B$-matrix &  $\begin{bmatrix}
                        ~0~&~0~\\
                        ~0~&~0~
                    \end{bmatrix}$
                 &  $\begin{bmatrix}
                        ~0~&~1~\\
                        ~1~&~0~
                    \end{bmatrix}$\\ \\[-0.5cm]
        \bottomrule 
    \end{tabular*}
    \caption{Unoriented graphs and the associated symmetric $B$-matrices for $n=2$ and $p=2$.}
    \label{tab:n2p2}
\end{table}

\paragraph{$B_{0}^{(2,2)}$ code}
The graph of the $B_{0}^{(2,2)}$ code is disconnected and the $B$-matrix is the zero-matrix.
The complete weight enumerator polynomial is the square of the one for the $B^{(1,2)}$ code, reflecting the disconnectedness of the graph:
\begin{align}
    W_{B_{0}^{(2,2)}}(\{x_{ab}\})
        =
            \left(W_{B^{(1,2)}}(\{x_{ab}\})\right)^2
         \ .
\end{align}
Hence, the partition function of the code CFT is given by
\begin{align}
        Z_{B_{0}^{(2,2)}}(\tau,\Bar{\tau}) 
            = \left( Z_{B^{(1,2)}}(\tau,\Bar{\tau}) \right)^2 \ ,
\end{align}
which shows that this CFT consists of two compact bosons of radius $R = 2$.

The $B_{0}^{(2,2)}$ code is non-$\BF_4$-even, and the code CFT has the non-anomalous $\BZ_2$ symmetry associated with the vector~\eqref{nonF4even_Z2}.
By gauging the $\BZ_2$ symmetry, one obtains the orbifolded theory with the partition function:
\begin{align}
    Z_{B_{0}^{(2,2)}}^{\CO}(\tau,\Bar{\tau})
        =
            \dfrac{|\theta_2|^4 + |\theta_3|^4 + |\theta_4|^4}{2\,|\eta(\tau)|^{4}} \ ,
\end{align}
where $\theta_i$ $(i=2,3,4)$ are the Jacobi theta functions defined by
\begin{align}
    \theta_2(\tau) = \sum_{n\in\BZ} q^{\frac{1}{2}(n-\frac{1}{2})^2}\,,\quad \theta_3(\tau) = \sum_{n\in\BZ} q^{\frac{n^2}{2}}\,,\quad \theta_4(\tau) = \sum_{n\in\BZ}(-1)^n q^{\frac{n^2}{2}}\,.
\end{align}
It follows that the orbifolded theory is the GSO projection of two copies of free Dirac fermions.
Similarly, the partition functions of the fermionized theory are given by
\begin{align}
    \begin{aligned}
        Z_{B_{0}^{(2,2)}}^{\widetilde{\mathrm{R}}}(\tau,\Bar{\tau})
            &=
                -\dfrac{|\theta_2|^4 + \left(|\theta_3|^2 - |\theta_4|^2\right)^2 - 2\,|\theta_2|^2 \left(|\theta_3|^2 + |\theta_4|^2\right)}{4\,|\eta(\tau)|^{4}} \ , \\
        Z_{B_{0}^{(2,2)}}^{\text{NS}}(\tau,\Bar{\tau})
            &=
                \dfrac{|\theta_2|^4 + \left(|\theta_3|^2 + |\theta_4|^2\right)^2 + 2\,|\theta_2|^2 \left(|\theta_3|^2 + |\theta_4|^2\right)}{4\,|\eta(\tau)|^{4}} \ , \\
        Z_{B_{0}^{(2,2)}}^{\text{R}}(\tau,\Bar{\tau})
            &=
                \dfrac{|\theta_2|^4 + \left(|\theta_3|^2 - |\theta_4|^2\right)^2 + 2\,|\theta_2|^2 \left(|\theta_3|^2 + |\theta_4|^2\right)}{4\,|\eta(\tau)|^{4}} \ .
    \end{aligned}
\end{align}
With these partition functions, we can check that the SUSY conditions are not met for the fermionized theory.

\paragraph{$B_{1}^{(2,2)}$ code}
Next, we turn to the $B_{1}^{(2,2)}$ code with the connected graph.
The complete weight enumerator polynomial is
\begin{align}
    W_{B_{1}^{(2,2)}}(\{x_{ab}\})
        =
            x_{00}^2 + 2\,x_{01} x_{10} + x_{11}^2 \ ,
\end{align}
and the partition function of the code CFT becomes
\begin{align}
        Z_{B_{1}^{(2,2)}}
            = 
                Z_{B_{0}^{(2,2)}}^\CO
            \ ,
\end{align}
which indicates that this CFT is the orbifolded theory of the $B_0^{(2,2)}$ code CFT.

The $B_{1}^{(2,2)}$ code is $\BF_4$-even, and the code CFT has the $\widehat\BZ_2$ symmetry associated with the vector \eqref{F4even_Z2}.
The $\widehat\BZ_2$ symmetry is, however, anomalous as $n=2 \not\in 4\,\BZ$, which prevents us from gauging the symmetry.
We can obtain the $B_0^{(2,2)}$ theory by orbifolding the $B_1^{(2,2)}$ theory with a vector $\mu\in\Lambda(B_1^{(2,2)})$ such that $\chi\circledcirc\mu=2$ mod $4$ and $\mu\circledcirc\mu\in8\BZ$ instead of $\chi$.
For example, $\mu = (1/\sqrt{2},4\sqrt{2},0,1/\sqrt{2})\in \Lambda(B_1^{(2,2)})$ where $c = (1,0,0,1)\in B_1^{(2,2)}$ and $m=(0,4,0,0)$ in the Construction A lattice \eqref{conA}.

\subsubsection{$p=3$}

For $p=3$, there are two oriented graphs as shown in Table \ref{tab:n2p3}.

\tikzstyle{node}=[draw=black,circle,inner sep=0,minimum size=10]
\begin{table}[t]
    \centering
    \renewcommand{\arraystretch}{1.5}
    \begin{tabular*}{10cm}{c@{\extracolsep{\fill}}cc}
        \toprule
       Name     & $B_{0}^{(2,3)}$ & $B_{1}^{(2,3)}$ \\\midrule \\[-0.4cm]
       Graph    &   \begin{tikzpicture}[scale=1.5, thick, >=stealth]
                        \node (v1) at (0,0) [node]{1} ;
                        \node (v2) at (1,0) [node]{2} ;
                    \end{tikzpicture}
                &   \begin{tikzpicture}[scale=1.5, thick, >=stealth]
                        \node (v1) at (0,0) [node]{1} ;
                        \node (v2) at (1,0) [node]{2} ;
                        \draw[->,>=stealth](v1)--(v2);
                    \end{tikzpicture} \\  [10pt] 
        $B$-matrix &  $\begin{bmatrix}
                        ~0~&~0~\\
                        ~0~&~0~
                    \end{bmatrix}$
                 &  $\begin{bmatrix}
                        ~0~&~1~\\
                        -1~&~0~
                    \end{bmatrix}$\\ \\[-0.5cm]
        \bottomrule 
    \end{tabular*}
    \caption{Oriented graphs and the associated anti-symmetric $B$-matrices for $n=2$ and $p=3$.}
    \label{tab:n2p3}
\end{table}

\paragraph{$B_{0}^{(2,3)}$ code}
Since the graph is disconnected, the complete weight enumerator polynomial becomes the square of the one for the $B^{(1,3)}$ code:
\begin{align}
    W_{B_{0}^{(2,3)}}(\{x_{ab}\})
        =
            \left(W_{B^{(1,3)}}(\{x_{ab}\})\right)^2
         \ .
\end{align}
Hence, the partition function of the code CFT is given by
\begin{align}
        Z_{B_{0}^{(2,3)}}
            = \left( Z_{B^{(1,3)}} \right)^2 \ ,
\end{align}
which shows that this CFT consists of two compact bosons of radius $R = \sqrt{6}$.

This theory has the non-anomalous $\BZ_2$ symmetry, which allows us to construct the orbifolded theory by gauging the symmetry.
The orbifold partition function is
\begin{align}
    \begin{aligned}
        Z_{B_{0}^{(2,3)}}^{\CO}
            =
                \dfrac{1}{|\eta|^{4}}&
                \left[ 1 + 4\,(q\bar q)^\frac{1}{6}+ 4\,(q\bar q)^\frac{1}{3} + 8\,(q\bar q)^\frac{5}{12} + 8\,q^\frac{13}{12}\,\bar q^\frac{1}{12} + 8\,q^\frac{1}{12}\,\bar q^\frac{13}{12} + 4\, (q\bar q)^\frac{2}{3} + 8\, (q\bar q)^\frac{3}{4} + \cdots \right] \ .
    \end{aligned}
\end{align}

The RR partition function of the fermionized theory is
\begin{align}
    \begin{aligned}
        Z_{B_{0}^{(2,3)}}^{\widetilde{\text{R}}}
            &=
                4 \ ,
    \end{aligned}
\end{align}
which is constant.
On the other hand, the $q$ expansions of the NS and R partition functions are
\begin{align}
    \begin{aligned}
        &Z_{B_{0}^{(2,3)}}^{\text{NS}}\\
            &\qquad =
                \dfrac{1}{|\eta|^{4}}\left[ 
                    1 + 4\,(q\bar q)^\frac{1}{6}+ 4\,(q\bar q)^\frac{1}{3} 
                    + 4\,q^\frac{2}{3}\,\bar q^\frac{1}{6} + 4\,q^\frac{1}{6}\,\bar q^\frac{2}{3}
                    + 8\,q^\frac{5}{6}\,\bar q^\frac{1}{3} + 8\,q^\frac{1}{3}\,\bar q^\frac{5}{6}
                    + 4\,(q\bar q)^\frac{2}{3} 
                    + 4\,q^\frac{3}{2} + 4\,\bar q^\frac{3}{2} + \cdots
                \right] \\
        &Z_{B_{0}^{(2,3)}}^{\text{R}}
            =     
                    4 + 16\,(q\bar q)^\frac{1}{3} + 16\,q + 16\,\bar q
                    + 16\,(q\bar q)^\frac{2}{3} 
                    + 48\,q^\frac{4}{3}\,\bar q^\frac{1}{3} + 48\,q^\frac{1}{3}\,\bar q^\frac{4}{3} + \cdots\ .
    \end{aligned}
\end{align}
Since the $\widetilde{\mathrm{R}}$ partition function is constant, the NS sector has dimension (anti-)chiral primary fields of dimension $3/2$, and the R sector satisfies the positive energy condition, $h, \bar h \ge n/24= 1/12$, this fermionic theory is expected to have supersymmetries.
Following the discussion in section~\ref{ss:fermion}, the theory has an actual $\CN=2$ supersymmetry, and the supercurrents are given by~\eqref{eq:N=2_susy}.

\paragraph{$B_{1}^{(2,3)}$ code}
The complete weight enumerator polynomial is given by
\begin{align}
    \begin{aligned}
    W_{B_{1}^{(2,3)}}(\{x_{ab}\})
        &=
            x_{00}^{2} + x_{01} x_{10} + x_{01} x_{20} + x_{02} x_{10} + x_{02} x_{20} \\
        &\qquad + 
            x_{11} x_{12} + x_{11} x_{21} + x_{12} x_{22} + x_{21} x_{22} \ .
    \end{aligned}
\end{align}
The partition function of the code CFT is
\begin{align}
    \begin{aligned}
        Z_{B_{1}^{(2,3)}}
            &=
            Z_{B_{0}^{(2,3)}}^{\CO} \ .
    \end{aligned}
\end{align}

The orbifold partition function by the (non-anomalous) $\BZ_2$ symmetry is
\begin{align}
    \begin{aligned}
        Z_{B_{1}^{(2,3)}}^{\CO}
            &=
                \dfrac{1}{|\eta(\tau)|^{2}}\left( 
                    \Theta_0\bar{\Theta}_0 + \Theta_1\bar{\Theta}_5 + \Theta_2\bar{\Theta}_4 + \Theta_3\bar{\Theta}_3 + \Theta_4\bar{\Theta}_2 + \Theta_5\bar{\Theta}_1  
                    \right)\cdot Z_{B^{(1,3)}} \\
            &=
                Z_{B_{0}^{(2,3)}}
                \ .
    \end{aligned}
\end{align}
Thus, the $B_1^{(2,3)}$ code CFT is the orbifolded theory of the $B_0^{(2,3)}$ code CFT.

The $\widetilde{\mathrm{R}}$ partition function of the fermionized theory is
\begin{align}
    \begin{aligned}
        Z_{B_{1}^{(2,3)}}^{\widetilde{\text{R}}}
            =
                -Z_{B_{0}^{(2,3)}}^{\widetilde{\text{R}}}
            =
                -4 \ ,
    \end{aligned}
\end{align}
which is constant.
On the other hand, the $q$ expansions of the NSNS and RNS partition functions take the same forms as those for the $B^{(2,3)}_0$ code:
\begin{align}
    \begin{aligned}
        Z_{B_{1}^{(2,3)}}^{\text{NS}}
            &=
            Z_{B_{0}^{(2,3)}}^{\text{NS}} \ , \\
        Z_{B_{1}^{(2,3)}}^{\text{R}}
            &=
            Z_{B_{0}^{(2,3)}}^{\text{R}}
            \ .
    \end{aligned}
\end{align}
Thus, the fermionized $B^{(2,3)}_0$ theory is equivalent to stacking the Kitaev Majorana chain with the fermionized $B^{(2,3)}_1$ theory~\cite{Ji:2019ugf}, and have the same number of supersymmetries.

\subsection{$n=3$}

\subsubsection{$p=2$}
For $p=2$, there are four inequivalent unoriented graphs, including the completely disconnected one, as shown in Table \ref{tab:n3p2}.

\begin{table}[t]
    \centering
    \renewcommand{\arraystretch}{1.5}
    \begin{tabular*}{15cm}{
    @{\extracolsep{\fill}}cccc}
        \toprule
       $B_{0}^{(3,2)}$ & $B_{1}^{(3,2)}$ & $B_{2}^{(3,2)}$ & $B_{3}^{(3,2)}$\\  \midrule\\[-0.4cm]
                \begin{tikzpicture}[scale=1, thick, baseline=-.5*(height("$+$")-depth("$+$")]
                          \foreach \i in{1,2,3}
                        \node (v\i) at ({cos((4*\i-1)*pi/6 r)},{sin((4*\i-1)*pi/6 r)}) [node]{} ;
                    \end{tikzpicture}
                &   \begin{tikzpicture}[scale=1, thick, baseline=-.5*(height("$+$")-depth("$+$")]
                          \foreach \i in{1,2,3}
                        \node (v\i) at ({cos((4*\i-1)*pi/6 r)},{sin((4*\i-1)*pi/6 r)}) [node]{} ;
                        \draw[-](v2)--(v3);
                    \end{tikzpicture}
                &   \begin{tikzpicture}[scale=1, thick, baseline=-.5*(height("$+$")-depth("$+$")]
                          \foreach \i in{1,2,3}
                        \node (v\i) at ({cos((4*\i-1)*pi/6 r)},{sin((4*\i-1)*pi/6 r)}) [node]{} ;
                        \draw[-](v1)--(v2);
                        \draw[-](v1)--(v3);
                    \end{tikzpicture}
                &   \begin{tikzpicture}[scale=1, thick, baseline=-.5*(height("$+$")-depth("$+$")]
                      \foreach \i in{1,2,3}
                        \node (v\i) at ({cos((4*\i-1)*pi/6 r)},{sin((4*\i-1)*pi/6 r)}) [node]{} ;
                        \draw[-](v1)--(v2);
                        \draw[-](v2)--(v3);
                        \draw[-](v1)--(v3);
                    \end{tikzpicture}
                \\  [30pt] 
        $\begin{bmatrix}
                            ~0~&~0~&~0~\\
                            ~0~&~0~&~0~\\
                            ~0~&~0~&~0~
                        \end{bmatrix}$
                    & $\begin{bmatrix}
                            ~0~&~1~&~0~\\
                            ~1~&~0~&~0~\\
                            ~0~&~0~&~0~
                        \end{bmatrix}$ 
                    & $\begin{bmatrix}
                        ~0~&~1~&~1~\\
                        ~1~&~0~&~0~\\
                        ~1~&~0~&~0~
                      \end{bmatrix}$
                    & $\begin{bmatrix}
                             ~0~&~1~&~1~\\
                            ~1~&~0~&~1~\\
                            ~1~&~1~&~0~
                        \end{bmatrix}$\\ \\[-0.5cm]
        \bottomrule 
    \end{tabular*}
    \caption{Unoriented graphs and the associated symmetric $B$-matrices for $n=3$ and $p=2$.}
    \label{tab:n3p2}
\end{table}

\paragraph{$B_{0}^{(3,2)}$ code}

The $B_{0}^{(3,2)}$ code corresponds to the completely disconnected graph, and the complete weight enumerator polynomial is given by the cube of the one for the $B^{(1,2)}$ code
\begin{align}
    W_{B_{0}^{(3,2)}}(\{x_{ab}\})
        =
            \left(W_{B^{(1,2)}}(\{x_{ab}\})\right)^3
         \ .
\end{align}
The code CFT consists of three copies of compact bosons of radius $R = 2$ as seen from the partition function:
\begin{align}
        Z_{B_{0}^{(3,2)}}(\tau,\Bar{\tau}) 
            = \left( Z_{B^{(1,2)}}(\tau,\Bar{\tau}) \right)^3 \ .
\end{align}

The $B_{0}^{(3,2)}$ code is non-$\BF_4$-even, and the code CFT has the non-anomalous $\BZ_2$ symmetry associated with the vector \eqref{nonF4even_Z2}.
By gauging the $\BZ_2$ symmetry, one obtains the orbifolded theory with the partition function:
\begin{align}
    Z_{B_{0}^{(3,2)}}^{\CO}(\tau,\Bar{\tau})
        =
            \dfrac{|\theta_2|^6 + 3\,|\theta_2|^2\left( |\theta_3|^2 - |\theta_4|^2\right)^2 + 3\,|\theta_2|^4\left( |\theta_3|^2 + |\theta_4|^2\right)  + \left( |\theta_3|^2 + |\theta_4|^2\right)^3}{8\,|\eta(\tau)|^{6}} \ .
\end{align}

While the original theory is just a product of three compact bosons of the same radius, its fermionic counterpart is a non-trivial fermionic CFT.
The $\widetilde{\mathrm{R}}$ partition function is
\begin{align}
    Z_{B_{0}^{(3,2)}}^{\widetilde{\mathrm{R}}}(\tau,\Bar{\tau})
        =
        \frac{3\,|\theta_2|^2\,|\theta_3|^2 \, |\theta_4|^2}{2\,|\eta(\tau)|^{6}}
        =
        6 \ ,
\end{align}
while the partition functions of the NS and R sectors are
\begin{align}
    \begin{aligned}
        Z_{B_{0}^{(3,2)}}^{\text{NS}}(\tau,\Bar{\tau})
            &=
                \frac{|\theta_3|^2\left(3\,|\theta_2|^4 + |\theta_3|^4 + 3\,|\theta_4|^4\right)}{4\,|\eta(\tau)|^{6}} \\
            &=
               \frac{1}{|\eta(\tau)|^{6}}
                \left[ 1 + 12\,(q\bar q)^\frac{1}{4} + 24\,q^\frac{3}{4}\bar q^\frac{1}{4} + 12\,(q\bar q)^\frac{1}{2} + 24\,q^\frac{1}{4}\bar q^\frac{3}{4} + 8\, q^\frac{3}{2} + 8\, \bar q^\frac{3}{2} \right.\\
                &\qquad\qquad\qquad\left. + 24\,q^\frac{5}{4}\bar q^\frac{1}{4} + 24\,q\bar q^\frac{1}{2} + 48\,(q\bar q)^\frac{3}{4} + 24\, q^\frac{1}{2}\bar q + 24\,q^\frac{1}{4}\bar q^\frac{5}{4} + 6\,q^2 + 6\bar q^2 + \cdots
                \right] \ , \\
        Z_{B_{0}^{(3,2)}}^{\text{R}}(\tau,\Bar{\tau})
            &=
                \dfrac{|\theta_2|^2\left( |\theta_2|^2 + 3\,|\theta_3|^2 +3\, |\theta_4|^2\right)^2}{4\,|\eta(\tau)|^{6}} \\
            &=
                6 + 16\,(q\bar q)^\frac{1}{4} + 48\,q + 96\,(q\bar q)^\frac{1}{2} + 48\,\bar q + 96\,q^\frac{5}{4}\bar q^\frac{1}{4} + 96\,q^\frac{1}{4}\bar q^\frac{5}{4} + 192\,q^2 + 192\bar q^2 + \cdots
                \ .
    \end{aligned}
\end{align}
We find that all the SUSY conditions are met and the fermionized CFT is expected to have supersymmetries.

\paragraph{$B_{1}^{(3,2)}$ code}

The $B_{1}^{(3,2)}$ code corresponds to the graph with one vertex and two vertices connected by a line.
Thus, the code CFT is a product of the $B^{(1,2)}$ code CFT (a compact boson of radius $R = \sqrt{6}$) and the $B_1^{(2,2)}$ code CFT:
\begin{align}
    Z_{B_{1}^{(3,2)}}
        =
        Z_{B^{(1,2)}}\cdot Z_{B_1^{(2,2)}} = Z_{B^{(1,2)}}\cdot Z_{B_0^{(2,2)}}^{\CO}\ .
\end{align}

The $B_{1}^{(3,2)}$ code is non-$\BF_4$-even and has the non-anomalous $\BZ_2$ symmetry.
The code CFT is self-dual under the gauging of the $\BZ_2$ symmetry:
\begin{align}
     Z_{B_{1}^{(3,2)}}^\CO
        =
         Z_{B_{1}^{(3,2)}} \ .
\end{align}
On the other hand, the fermionic partition functions are
\begin{align}
    \begin{aligned}
        Z_{B_{1}^{(3,2)}}^{\widetilde{\mathrm{R}}}
            &=
                0 \ , \\
        Z_{B_{1}^{(3,2)}}^\text{NS}
            &=
                \dfrac{|\theta_3(\tau)|^2}{|\eta(\tau)|^{2}} \cdot Z_{B_{1}^{(2,2)}} \ , \\
        Z_{B_{1}^{(3,2)}}^\text{R}
            &=
                 \dfrac{|\theta_2(\tau)|^2}{|\eta(\tau)|^{2}} \cdot Z_{B_{1}^{(2,2)}} \ , 
    \end{aligned}
\end{align}
which shows that the fermionized CFT is a product of one free Dirac fermion and the $B_{1}^{(2,2)}$ code CFT, which is the GSO projection of two copies of free Dirac fermions.
One can check that all the SUSY conditions are met, implying that the fermionized theory has supersymmetries.
The partition function $Z_{B_{1}^{(3,2)}}^\text{R}$ in the $\text{R}$ sector, however, does not have a constant in the $q$ expansion, hence if the theory is supersymmetric, it has to be spontaneously broken.

\paragraph{$B_{2}^{(3,2)}$ code}
The partition function of $B_{2}^{(3,2)}$ code CFT turns out to be the one of the orbifold of the $B_{0}^{(3,2)}$ code CFT:
\begin{align}
    Z_{B_{2}^{(3,2)}}(\tau, \bar\tau)
        =
        Z_{B_0^{(3,2)}}^\CO(\tau, \bar\tau) \ .
\end{align}
Since the $B_{2}^{(3,2)}$ code is non-$\BF_4$-even and has the non-anomalous $\BZ_2$ symmetry, which leads us to the $B_{0}^{(3,2)}$ code CFT by the orbifolding:
\begin{align}
     Z_{B_{2}^{(3,2)}}^\CO(\tau, \bar\tau) 
        =
         Z_{B_{0}^{(3,2)}}(\tau, \bar\tau) \ .
\end{align}
The fermionized theory is equivalent to stacking the Kitaev-Majorana chain with the fermionic $B_{0}^{(3,2)}$ code CFT and has the same supersymmetries as the latter.

\paragraph{$B_{3}^{(3,2)}$ code}
The $B_{3}^{(3,2)}$ code is non-$\BF_4$-even and the corresponding CFT has the partition function
\begin{align}
    Z_{B_{3}^{(3,2)}}(\tau, \bar \tau)
        =
        \dfrac{|\theta_2|^6 + |\theta_3|^6 + |\theta_4|^6}{2\,|\eta(\tau)|^{6}} \ .
\end{align}
This is the GSO projection of three copies of free Dirac fermions.
This theory has the non-anomalous $\BZ_2$ symmetry, under which it is self-dual by orbifolding:
\begin{align}
    Z_{B_{3}^{(3,2)}}^\CO(\tau, \bar\tau)
        =
        Z_{B_3^{(3,2)}}(\tau, \bar\tau) \ .
\end{align}
The fermionized theory is a product of three copies of free Dirac fermions as seen from the fermionic partition functions:
\begin{align}
    \begin{aligned}
        Z_{B_{3}^{(3,2)}}^{\widetilde{\mathrm{R}}}(\tau, \bar\tau) 
            &=
                0 \ , \\
        Z_{B_{3}^{(3,2)}}^\text{NS}(\tau, \bar\tau) 
            &=
                \dfrac{|\theta_3(\tau)|^6}{|\eta(\tau)|^{6}} \ , \\
        Z_{B_{3}^{(3,2)}}^\text{R}(\tau, \bar\tau) 
            &=
                \dfrac{|\theta_2(\tau)|^6}{|\eta(\tau)|^{6}} \ .
    \end{aligned}
\end{align}

\subsubsection{$p=3$}
\begin{table}[th]
    \centering
    \renewcommand{\arraystretch}{1.5}
    \begin{tabular*}{\textwidth}{
    @{\extracolsep{\fill}}cccccc}
        \toprule
       $B_{1}^{(3,3)}$ & $B_{2}^{(3,3)}$ & $B_{3}^{(3,3)}$ & $B_{4}^{(3,3)}$ & $B_{5}^{(3,3)}$  & $B_{6}^{(3,3)}$ \\  \midrule\\[-0.4cm]
       \begin{tikzpicture}[scale=0.8, thick, baseline=-.5*(height("$+$")-depth("$+$")]
                          \foreach \i in{1,2,3}
                        \node (v\i) at ({cos((4*\i-1)*pi/6 r)},{sin((4*\i-1)*pi/6 r)}) [node]{\i} ;
                        \draw[->,>=stealth](v2)--(v3);
                    \end{tikzpicture}
                &   \begin{tikzpicture}[scale=0.8, thick, baseline=-.5*(height("$+$")-depth("$+$")]
                          \foreach \i in{1,2,3}
                        \node (v\i) at ({cos((4*\i-1)*pi/6 r)},{sin((4*\i-1)*pi/6 r)}) [node]{\i} ;
                        \draw[->,>=stealth](v1)--(v2);
                        \draw[->,>=stealth](v3)--(v2);
                    \end{tikzpicture}
                &   \begin{tikzpicture}[scale=0.8, thick, baseline=-.5*(height("$+$")-depth("$+$")]
                          \foreach \i in{1,2,3}
                        \node (v\i) at ({cos((4*\i-1)*pi/6 r)},{sin((4*\i-1)*pi/6 r)}) [node]{\i} ;
                        \draw[->,>=stealth](v1)--(v2);
                        \draw[->,>=stealth](v2)--(v3);
                    \end{tikzpicture}

                &   \begin{tikzpicture}[scale=0.8, thick, baseline=-.5*(height("$+$")-depth("$+$")]
                          \foreach \i in{1,2,3}
                        \node (v\i) at ({cos((4*\i-1)*pi/6 r)},{sin((4*\i-1)*pi/6 r)}) [node]{\i} ;
                        \draw[->,>=stealth](v2)--(v1);
                        \draw[->,>=stealth](v2)--(v3);
                    \end{tikzpicture}

                &   \begin{tikzpicture}[scale=0.8, thick, baseline=-.5*(height("$+$")-depth("$+$")]
                          \foreach \i in{1,2,3}
                        \node (v\i) at ({cos((4*\i-1)*pi/6 r)},{sin((4*\i-1)*pi/6 r)}) [node]{\i} ;
                        \draw[->,>=stealth](v1)--(v2);
                        \draw[->,>=stealth](v2)--(v3);
                        \draw[->,>=stealth](v3)--(v1);
                    \end{tikzpicture}
                &   \begin{tikzpicture}[scale=0.8, thick, baseline=-.5*(height("$+$")-depth("$+$")]
                      \foreach \i in{1,2,3}
                        \node (v\i) at ({cos((4*\i-1)*pi/6 r)},{sin((4*\i-1)*pi/6 r)}) [node]{\i} ;
                        \draw[->,>=stealth](v1)--(v2);
                        \draw[->,>=stealth](v2)--(v3);
                        \draw[->,>=stealth](v1)--(v3);
                    \end{tikzpicture}
                \\  [30pt] 
                {\footnotesize
                    $\begin{bmatrix}
                            ~0~&~0~&~0~\\
                            ~0~&~0~&~1\\
                            ~0~&-1~&~0~
                        \end{bmatrix}$
                }
                    &
                {\footnotesize
                    $\begin{bmatrix}
                            ~0~&~1~&~0~\\
                            -1~&~0~&-1~\\
                            ~0~&~1~&~0~
                        \end{bmatrix}$ 
                }
                    &
                {\footnotesize
                    $\begin{bmatrix}
                            ~0~&~1~&~0~\\
                            -1~&~0~&~1~\\
                            ~0~&-1~&~0~
                        \end{bmatrix}$
                }
                    & 
                {\footnotesize
                    $\begin{bmatrix}
                            ~0~&-1~&~0~\\
                            ~1~&~0~&~1~\\
                            ~0~&-1~&~0~
                      \end{bmatrix}$
                }
                    & 
                {\footnotesize
                    $\begin{bmatrix}
                        ~0~&~1~&-1~\\
                        -1~&~0~&~1~\\
                        ~1~&-1~&~0~
                      \end{bmatrix}$
                }
                    & 
                {\footnotesize
                    $\begin{bmatrix}
                             ~0~&~1~&~1~\\
                            -1~&~0~&~1~\\
                            -1~&-1~&~0~
                        \end{bmatrix}$
                }\\ \\[-0.5cm]
        \bottomrule 
    \end{tabular*}
    \caption{Oriented graphs and the associated anti-symmetric $B$-matrices for $n=3$ and $p=3$.}
    \label{tab:n3p3}
\end{table}

For $n=3$, there are six non-equivalent oriented graphs.
The adjacency matrices corresponding to these graphs are given in Table \ref{tab:n3p3}.

The graph for $B^{(3,3)}_1$ consists of one vertex graph $B^{(1,3)}$ and two vertex graph $B^{(2,3)}_1$.
Thus, the weight enumerator polynomial factorizes, and the resulting code CFT is a product of a single compact boson of radius $R=\sqrt{6}$ and the B-form code CFT of $B^{(2,3)}_1$ type.

$B^{(3,3)}_5$ corresponds to the Paley matrix $Q_3$, so the classical code with the generator matrix $\SH_{\text{sub}}=\left[\,I\,\Big|\,B^{(3,3)}_5\,\right]$ is one of the quadratic double circulant codes $\CP_{6}$.
By using $W_{B^{(3,3)}_5}$, the partition function of the code CFT for $B^{(3,3)}_5$ is
\begin{align}
    \begin{aligned}
        Z_{B^{(3,3)}_5}(\tau,\Bar{\tau})
                
        &=
           Z_{B^{(1,3)}}(\tau, \bar\tau) \cdot Z_\text{remain}(\tau, \bar\tau) \ ,
    \end{aligned}
\end{align}
where 
\begin{align}
    Z_\text{remain}(\tau, \bar\tau)
        =
            \dfrac{1}{|\eta(\tau)|^{4}}\left[ \left(\sum_{l=0}^5 |\Theta_l|^2\right)^2
                - 3\sum_{l=0}^2 |\Theta_l \Theta_{l+1} - \Theta_{l+3}\Theta_{l+4}|^2 \right] \ .
\end{align}
Since $Z_\text{remain}$ is modular invariant by itself, one may expect that the $B^{(3,3)}_5$ theory is a product of the $B^{(1,3)}$ code CFT and a CFT with $c=2$.
It is, however, checked by the $q$ expansion that $Z_\text{remain}$ cannot describe a CFT with $c=2$ as the expansion includes negative coefficients:
\begin{align}
    Z_\text{remain}
        =
        \frac{1}{|\eta|^4}\left[ 1
        -2\, (q \bar q)^{\frac{1}{12}} + 4\,(q \bar q)^{\frac{1}{6}} + 4\,(q \bar q)^{\frac{1}{3}} + 8\,(q \bar q)^{\frac{5}{12}} + 12\,q^\frac{13}{12}\bar q^\frac{1}{12} + 12\,q^\frac{1}{12}\bar q^\frac{13}{12} + \cdots \right] \ .
\end{align}
Hence, the $B^{(3,3)}_5$ theory is a CFT with $c=3$, which does not factorize into two CFTs.
We note that the partition function $Z_{B^{(3,3)}_5}$ has the expansion with positive coefficients:
\begin{align}
    Z_{B^{(3,3)}_5}
            =
            \frac{1}{|\eta|^6}\left[ 1 + 8\,(q \bar q)^{\frac{1}{4}} + 6\,(q \bar q)^{\frac{1}{3}} + 12\,(q\bar q)^{\frac{5}{12}} + 24\, (q\bar q)^\frac{1}{2} + 12\,q^\frac{13}{12}\bar q^\frac{1}{12} + 12\,q^\frac{1}{12}\bar q^\frac{13}{12} + \cdots \right]  \ .
\end{align}
Since there are no terms proportional to $q$ and $\bar q$ in the square bracket, we find that this theory does not have current operators constructed as vertex operators.
Thus, this theory cannot be described by WZW models of simple Lie algebras.

While all the code CFTs listed above have the $\BZ_2$ symmetries, they are anomalous and cannot be gauged to construct the orbifolded and fermionized theories.

\subsection{$n=4, p=2$}

We select some of the B-form codes of length $n=4$ over $\BF_2$, which have special properties.

\paragraph{$B_1^{(4,2)}$ code}
First, we consider the following $B_1^{(4,2)}$ code:
\begin{center}
    \begin{tikzpicture}[scale=2, thick]
        \node at (0,0) {\large $B_1^{(4,2)}$\,:};
        \begin{scope}[xshift=2cm,scale=0.7]
            \foreach \i in{1,...,4}{
            \node (v\i) at ({cos((3*pi/4+2*(\i-1)*pi/4) r)},{sin((3*pi/4+2*(\i-1)*pi/4) r)}) [node]{} ;}
    
             \draw[-](v1)--(v2);
             \draw[-](v1)--(v3);
             \draw[-](v1)--(v4);
        \end{scope}

         \begin{scope}[xshift=4cm]
            \node[right] at (0,0) {
                $B = \begin{bmatrix}
                        ~0~ & ~1~ & ~1~ & ~1~ \\
                        ~1~ & ~0~ & ~0~ & ~0~ \\
                        ~1~ & ~0~ & ~0~ & ~0~ \\
                        ~1~ & ~0~ & ~0~ & ~0~ \\
                     \end{bmatrix}$};
         \end{scope}
    \end{tikzpicture}
\end{center}
The partition function reads
\begin{align}
    Z_{B_1^{(4,2)}}(\tau, \bar\tau)
        &=
        \frac{1}{8\,|\eta(\tau)|^8}\left[|\theta_2|^8 + |\theta_3|^8 + |\theta_4|^8 + 6\left( |\theta_2|^4\,|\theta_3|^4 + |\theta_2|^4\,|\theta_4|^4 + |\theta_3|^4\,|\theta_4|^4\right) \right] \ .
\end{align}
This code is $\BF_4$-even, and the code CFT has the $\widehat\BZ_2$ symmetry.
By gauging the $\widehat\BZ_2$ symmetry, we obtain the orbifold partition function:
\begin{align}
    Z_{B_1^{(4,2)}}^\CO(\tau, \bar\tau)
        =
        Z_{B_1^{(4,2)}}(\tau, \bar\tau) \ ,
\end{align}
which shows that the $B_1^{(4,2)}$ code CFT is self-dual under the $\widehat\BZ_2$ symmetry.

The partition functions of the fermionized theory become
\begin{align}
    \begin{aligned}
        Z_{B_1^{(4,2)}}^{\widetilde{\mathrm{R}}}(\tau, \bar\tau) 
            &=
             0 \ , \\
        Z_{B_1^{(4,2)}}^\text{NS}(\tau, \bar\tau) 
            &=
             \frac{|\theta_3|^2}{|\eta|^2}\cdot Z_{B_0^{(3,2)}}^\text{NS} \ , \\
        Z_{B_1^{(4,2)}}^\text{R}(\tau, \bar\tau) 
            &=
             \frac{|\theta_2|^2}{|\eta|^2}\cdot Z_{B_0^{(3,2)}}^\text{R}  \ ,
    \end{aligned}
\end{align}
which indicates that the fermionic CFT is a product of a free Dirac fermion and the fermionized theory of the $B_0^{(3,2)}$ code CFT.
This theory satisfies the SUSY conditions and is expected to have spontaneously broken supersymmetries.

\paragraph{$B_2^{(4,2)}$ code}
Next, we consider the following $B_2^{(4,2)}$ code:
\begin{center}
    \begin{tikzpicture}[scale=2, thick]
         \node at (0,0) {\large $B_2^{(4,2)}$\,:};
        \begin{scope}[xshift=2cm,scale=0.7]
            \foreach \i in{1,...,4}{
            \node (v\i) at ({cos((3*pi/4+2*(\i-1)*pi/4) r)},{sin((3*pi/4+2*(\i-1)*pi/4) r)}) [node]{} ;}

            \foreach [count=\r] \row in {{0, 1, 1, 1}, {-1, 0, 1, -1}, {-1, -1, 0, 1}, {-1, 1, -1, 0}}{
                \foreach [count=\c] \cell in \row{
                    \ifnum\cell=1
                        \draw[-](v\r)--(v\c);
                    \fi
                }
                }
        \end{scope}
        
         \begin{scope}[xshift=4cm]
            \node[right] at (0,0) {
                $B = \begin{bmatrix}
                        ~0~ & ~1~ & ~1~ & ~1~ \\
                        ~1~ & ~0~ & ~1~ & ~1~ \\
                        ~1~ & ~1~ & ~0~ & ~1~ \\
                        ~1~ & ~1~ & ~1~ & ~0~ \\
                     \end{bmatrix}$};
         \end{scope}
    \end{tikzpicture}
\end{center}
This code is also $\BF_4$-even, and the code CFT has the $\widehat\BZ_2$ symmetry.

The partition function of the code CFT is
\begin{align}
    Z_{B_2^{(4,2)}}(\tau, \bar\tau)
        &=
        \frac{|\theta_2|^8 + |\theta_3|^8 + |\theta_4|^8}{2\,|\eta(\tau)|^8} \ .
\end{align}
This is the GSO projection of four copies of free Dirac fermions.
By gauging the $\widehat\BZ_2$ symmetry, we obtain the orbifold partition function:
\begin{align}
    Z_{B_2^{(4,2)}}^\CO(\tau, \bar\tau)
        =
        Z_{B_2^{(4,2)}}(\tau, \bar\tau) \ ,
\end{align}
which shows that the $B_2^{(4,2)}$ code CFT is also self-dual under the $\widehat\BZ_2$ symmetry.

The partition functions of the fermionized theory read
\begin{align}
    \begin{aligned}
        Z_{B_2^{(4,2)}}^{\widetilde{\mathrm{R}}}(\tau, \bar\tau) 
            &=
             0 \ , \\
        Z_{B_2^{(4,2)}}^\text{NS}(\tau, \bar\tau) 
            &=
             \frac{|\theta_3|^8}{|\eta|^8} \ , \\
        Z_{B_2^{(4,2)}}^\text{R}(\tau, \bar\tau) 
            &=
            \frac{|\theta_2|^8}{|\eta|^8}  \ ,
    \end{aligned}
\end{align}
which shows that the fermionic CFT is a product of four copies of free Dirac fermions.
The theory can be described in terms of 8 Majorana fermions and enjoys the triality, which exchanges the vector representation and the two spinor representation (see~\cite{Karch:2019lnn} for a recent discussion).

\paragraph{$B_3^{(4,2)}$ code}
Finally, consider the $B_3^{(4,2)}$ code whose corresponding graph is disconnected:
\begin{center}
    \begin{tikzpicture}[scale=2, thick]
         \node at (0,0) {\large $B_3^{(4,2)}$\,:};
        \begin{scope}[xshift=2cm,scale=0.7]
            \foreach \i in{1,...,4}{
            \node (v\i) at ({cos((3*pi/4+2*(\i-1)*pi/4) r)},{sin((3*pi/4+2*(\i-1)*pi/4) r)}) [node]{} ;}

            \draw[-](v1)--(v4);
            \draw[-](v2)--(v3);
        \end{scope}
        
         \begin{scope}[xshift=4cm]
            \node[right] at (0,0) {
                $B = \begin{bmatrix}
                        ~0~ & ~1~ & ~0~ & ~0~ \\
                        ~1~ & ~0~ & ~0~ & ~0~ \\
                        ~0~ & ~0~ & ~0~ & ~1~ \\
                        ~0~ & ~0~ & ~1~ & ~0~ \\
                     \end{bmatrix}$};
         \end{scope}
    \end{tikzpicture}
\end{center}
This code is also $\BF_4$-even, and the code CFT has the $\widehat\BZ_2$ symmetry.

Since the graph consists of two disconnected graphs of $B_1^{(2,2)}$ type, the partition function of the code CFT becomes
\begin{align}
    Z_{B_3^{(4,2)}}(\tau, \bar\tau)
        &=
        \left(Z_{B_1^{(2,2)}}(\tau, \bar\tau)\right)^2 \ .
\end{align}
This theory is also self-dual under the $\widehat\BZ_2$ symmetry as seen from the orbifold partition function:
\begin{align}
    Z_{B_3^{(4,2)}}^\CO(\tau, \bar\tau)
        =
        Z_{B_3^{(4,2)}}(\tau, \bar\tau) \ .
\end{align}

On the other hand, the partition functions of the fermionized theory are 
\begin{align}
    \begin{aligned}
        Z_{B_2^{(4,2)}}^{\widetilde{\mathrm{R}}}(\tau, \bar\tau) 
            &=
             0 \ , \\
        Z_{B_2^{(4,2)}}^\text{NS}(\tau, \bar\tau) 
            &=
             \frac{|\theta_3|^4}{|\eta|^4}\cdot Z_{B_1^{(2,2)}} \ , \\
        Z_{B_2^{(4,2)}}^\text{R}(\tau, \bar\tau) 
            &=
            \frac{|\theta_2|^8}{|\eta|^4}\cdot Z_{B_1^{(2,2)}}   \ ,
    \end{aligned}
\end{align}
which shows that the fermionic CFT is a product of two free Dirac fermions and the (bosonic) $B_1^{(2,2)}$ code CFT.
The fermionized theory satisfies the SUSY conditions and is expected to have spontaneously broken supersymmetries.

\subsection{$n=5,~ p=2$}
While there are many B-form codes of length $n=5$, we focus on the following $B_1^{(5,2)}$ code corresponding to the pentagon graph.
\begin{center}
    \begin{tikzpicture}[scale=2, thick]
        \node at (0,0) {\large $B_1^{(5,2)}$\,:};
        \begin{scope}[xshift=2cm,scale=0.8]
            \foreach \i in{1,...,5}{
            \node (v\i) at ({cos((pi/2+2*(\i-1)*pi/5) r)},{sin((pi/2+2*(\i-1)*pi/5) r)}) [node]{} ;}
    
             \draw[-](v1)--(v2);
             \draw[-](v2)--(v3);
             \draw[-](v3)--(v4);
             \draw[-](v4)--(v5);
             \draw[-](v5)--(v1);
        \end{scope}

         \begin{scope}[xshift=4cm]
            \node[right] at (0,0) {
                $B = \begin{bmatrix}
                        ~0~ & ~1~ & ~0~ & ~0~ & ~1~ \\
                        ~1~ & ~0~ & ~1~ & ~0~ & ~0~ \\
                        ~0~ & ~1~ & ~0~ & ~1~ & ~0~ \\
                        ~0~ & ~0~ & ~1~ & ~0~ & ~1~ \\
                        ~1~ & ~0~ & ~0~ & ~1~ & ~0~ \\
                     \end{bmatrix}$};
         \end{scope}
    \end{tikzpicture}
\end{center}
The code CFT is non-$\BF_4$-even and has the $\BZ_2$ symmetry.
The partition functions of the code CFT and its orbifolding are
\begin{align}
    \begin{aligned}
        Z_{B_1^{(5,2)}}
            &=
            Z_{B_1^{(5,2)}}^\CO \\
            &=
            \frac{1}{32\,|\eta|^{10}}\,\left( |\theta_2|^2 + |\theta_3|^2 + |\theta_4|^2\right)^2\, 
            \left[ \left( |\theta_2|^2 + |\theta_3|^2 + |\theta_4|^2\right)^3 -20\,|\theta_2|^2\, |\theta_3|^2\, |\theta_4|^2 \right] \ .
    \end{aligned}
\end{align}
Thus, the theory is self-dual under the $\BZ_2$-gauging.

The partition functions of the fermionized theory are given by
\begin{align}
    \begin{aligned}
        Z_{B_1^{(5,2)}}^{\widetilde{\mathrm{R}}}
            &=
             0 \ , \\
        Z_{B_1^{(5,2)}}^\text{NS}
            &=
             \frac{1}{16\,|\eta|^{10}}\, |\theta_3|^2 \left[ |\theta_3|^8 + 5\left( |\theta_2|^4 - |\theta_4|^4\right)^2 + 10\,|\theta_3|^4\left( |\theta_2|^4 + |\theta_4|^4\right)\right]\ , \\
             &=
             \frac{1}{|\eta|^{10}}\, \left[  1 + 80\,\left(q^\frac{3}{4}\bar q^\frac{1}{4} + q^\frac{1}{4}\bar q^\frac{3}{4}\right) + 100\,(q\bar q)^\frac{1}{2} \right. \\
             & \qquad\qquad \qquad\quad \left. + 160\,\left(q^\frac{5}{4}\bar q^\frac{1}{4} + \,q\bar q^\frac{1}{2} + 2\,(q\bar q)^\frac{3}{4} + q^\frac{1}{2}\bar q + q^\frac{1}{4}\bar q^\frac{5}{4}\right) + \cdots\right] \ ,\\
        Z_{B_1^{(5,2)}}^\text{R}
            &=
            \frac{1}{16\,|\eta|^{10}} \,|\theta_2|^2\left[ |\theta_2|^8 + 5\left( |\theta_3|^4 - |\theta_4|^4\right)^2 + 10\,|\theta_2|^4\left( |\theta_3|^4 + |\theta_4|^4\right)\right] \\
            &=
            80\,(q\bar q)^\frac{1}{6} + 80\,q^\frac{11}{12}\bar q^{-\frac{1}{12}} + 80\,q^{-\frac{1}{12}}\bar q^{\frac{11}{12}}+  \cdots 
            \ .
    \end{aligned}
\end{align}
While the $\widetilde{\mathrm{R}}$ partition function is vanishing and constant, the other SUSY conditions are not met, and the fermionized theory is not supersymmetric. 

The fermionic partition functions factorize to those of a free Dirac fermion and the modular invariant functions.
We note, however, that the modular invariant functions cannot be fermionic partition functions of a $c=4$ CFT as they have negative coefficients in the $q$ expansion.
Hence, the fermionized theory is a fermionic CFT that does not contain a free Dirac fermion.

\subsection{$n=6,~ p=3$}
Among various B-form code of length $n=6$, we consider the $B_1^{(6,3)}$ code corresponding to the following graph:
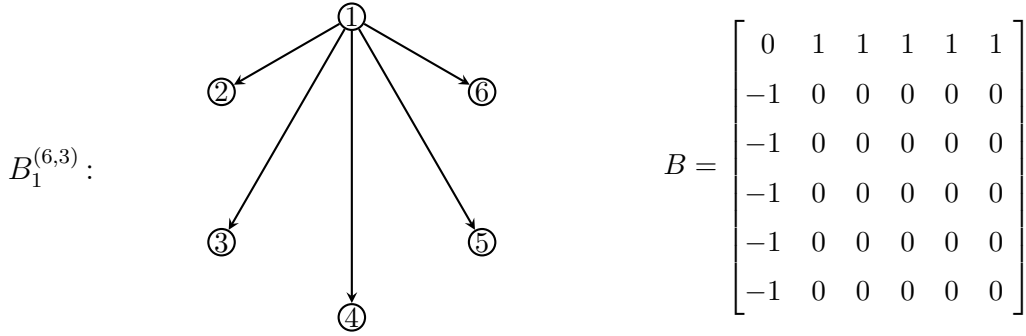
\begin{center}
    \begin{tikzpicture}[scale=2, thick]
        \node at (0,0) {\large $B_1^{(6,3)}$\,:};
        \begin{scope}[xshift=2cm]
            \foreach \i in{1,...,6}{
            \node (v\i) at ({cos((pi/2+2*(\i-1)*pi/6) r)},{sin((pi/2+2*(\i-1)*pi/6) r)}) [node]{\i} ;}
            
                \foreach [count=\r] \row in {
                    {0, 1, 1, 1, 1, 1},
                    {-1, 0, 0, 0, 0, 0}, 
                    {-1, 0, 0, 0, 0, 0}, 
                    {-1, 0, 0, 0, 0, 0}, 
                    {-1, 0, 0, 0, 0, 0}, 
                    {-1, 0, 0, 0, 0, 0}}{
                    \foreach [count=\c] \cell in \row{
                        \ifnum\cell=1
                            \draw[->,>=stealth](v\r)--(v\c);
                        \fi
                    }
                    }
        \end{scope}

         \begin{scope}[xshift=4cm]
            \node[right] at (0,0) {
                $B = \begin{bmatrix}
                        ~0~ & ~1~ & ~1~ & ~1~ & ~1~ & ~1~ \\
                        -1~ & ~0~ & ~0~ & ~0~ & ~0~ & ~0~ \\
                        -1~ & ~0~ & ~0~ & ~0~ & ~0~ & ~0~ \\
                        -1~ & ~0~ & ~0~ & ~0~ & ~0~ & ~0~ \\
                        -1~ & ~0~ & ~0~ & ~0~ & ~0~ & ~0~ \\
                        -1~ & ~0~ & ~0~ & ~0~ & ~0~ & ~0~ \\
                     \end{bmatrix}$};
         \end{scope}
    \end{tikzpicture}
\end{center}
The partition function of the corresponding CFT is 
\begin{align}
    Z_{B_1^{(6,3)}}
        =
        \frac{1}{|\eta|^{12}}\left[ 1 + 30\,(q\bar q)^\frac{1}{6} + 40\,(q\bar q)^\frac{1}{4} + 90\,(q\bar q)^\frac{1}{3} + 120\,(q\bar q)^\frac{5}{12} + 144\,(q\bar q)^\frac{1}{2} + 360\,(q\bar q)^\frac{7}{12} + \cdots \right] \ .
\end{align}

The $B_1^{(6,3)}$ code has the non-anomalous $\BZ_2$ symmetry.
The partition functions of the orbifolded and fermionized theories are given by
\begin{align}
    \begin{aligned}
        Z_{B_1^{(6,3)}}^\CO
            &=
                Z_{B_1^{(6,3)}} \ ,\\
        Z_{B_1^{(6,3)}}^{\widetilde{\mathrm{R}}}
            &=
                0 \ , \\
        Z_{B_1^{(6,3)}}^\text{NS}
            &= 
                \frac{1}{|\eta|^{12}}
                    \left[ 1 + 30\,(q\bar q)^\frac{1}{6} + 2\,q^\frac{1}{2} + 2\,\bar q^\frac{1}{2} + 90\,(q\bar q)^\frac{1}{3} + 60\,q^\frac{2}{3}\bar q^\frac{1}{6} + 60\,q^\frac{1}{6}\bar q^\frac{2}{3} \right. \\
                    &\left.\qquad \qquad \quad + 144\,(q\bar q)^\frac{1}{2} + 180\,q^\frac{5}{6}\bar q^\frac{1}{3} + 180\,q^\frac{1}{3}\bar q^\frac{5}{6} + 390\,(q\bar q)^\frac{2}{3} + 400\,q\bar q^\frac{1}{2} + 400\,q^\frac{1}{2}\bar q +  \cdots \right] \ , \\
        Z_{B_1^{(6,3)}}^\text{R}
            &= 
             80 + 240\,(q\bar q)^\frac{1}{6} + 720\,q^\frac{1}{3} + 560\,q + 576\,(q\bar q)^\frac{1}{2} + 560\,\bar q + \cdots \ .
    \end{aligned}
\end{align}
It follows that the code CFT is self-dual under the $\BZ_2$ symmetry, and the fermionized theory by the $\BZ_2$ symmetry does not have any supersymmetry.

\subsection{$n=12,\, p=2$}
We consider the B-form code of length 12 corresponding to the all-to-all graph:

\begin{center}
\begin{tikzpicture}
\node at (0,0) {\large $B_1^{(12,2)}$\,:};
\begin{scope}[scale=3,xshift=1.5cm]
    \foreach \i in {1,...,12}
    {
    \node[circle, draw, fill=white, inner sep=1.5pt] (n\i) at ({360/12 * (\i - 1)}:1) {};}

    \foreach \i in {1,...,12}
    {
        \foreach \j in {\i,...,12}
        {
            \ifnum \i < \j
                \draw (n\i) -- (n\j);
            \fi
        }
    }
    \end{scope}
    \begin{scope}[xshift=8cm]
            \node[right] at (0,0) {\footnotesize
                $B = \begin{bmatrix}
                        ~0~ & ~1~ & ~1~ & ~1~ & ~1~ & ~1~ & ~1~ & ~1~ & ~1~ & ~1~ & ~1~ & ~1~ \\
 ~1~ & ~0~ & ~1~ & ~1~ & ~1~ & ~1~ & ~1~ & ~1~ & ~1~ & ~1~ & ~1~ & ~1~ \\
 ~1~ & ~1~ & ~0~ & ~1~ & ~1~ & ~1~ & ~1~ & ~1~ & ~1~ & ~1~ & ~1~ & ~1~ \\
 ~1~ & ~1~ & ~1~ & ~0~ & ~1~ & ~1~ & ~1~ & ~1~ & ~1~ & ~1~ & ~1~ & ~1~ \\
 ~1~ & ~1~ & ~1~ & ~1~ & ~0~ & ~1~ & ~1~ & ~1~ & ~1~ & ~1~ & ~1~ & ~1~ \\
 ~1~ & ~1~ & ~1~ & ~1~ & ~1~ & ~0~ & ~1~ & ~1~ & ~1~ & ~1~ & ~1~ & ~1~ \\
 ~1~ & ~1~ & ~1~ & ~1~ & ~1~ & ~1~ & ~0~ & ~1~ & ~1~ & ~1~ & ~1~ & ~1~ \\
 ~1~ & ~1~ & ~1~ & ~1~ & ~1~ & ~1~ & ~1~ & ~0~ & ~1~ & ~1~ & ~1~ & ~1~ \\
 ~1~ & ~1~ & ~1~ & ~1~ & ~1~ & ~1~ & ~1~ & ~1~ & ~0~ & ~1~ & ~1~ & ~1~ \\
 ~1~ & ~1~ & ~1~ & ~1~ & ~1~ & ~1~ & ~1~ & ~1~ & ~1~ & ~0~ & ~1~ & ~1~ \\
 ~1~ & ~1~ & ~1~ & ~1~ & ~1~ & ~1~ & ~1~ & ~1~ & ~1~ & ~1~ & ~0~ & ~1~ \\
 ~1~ & ~1~ & ~1~ & ~1~ & ~1~ & ~1~ & ~1~ & ~1~ & ~1~ & ~1~ & ~1~ & ~0~ \\
                     \end{bmatrix}$};
         \end{scope}
\end{tikzpicture}
\end{center}

The complete weight enumerator is given by
\begin{align}
\begin{aligned}
    W_{B_1^{(12,2)}} = x_{0 0}^{12} &+ 66\,x_{0 0}^{10}x_{1 1}^2 + 495\,x_{0 0}^8x_{1 1}^4 + 924\,x_{0 0}^6x_{1 1}^6 + 495\,x_{0 0}^4x_{1 1}^8 + 66\,x_{0 0}^2x_{1 1}^{10} + 12\,x_{0 1}^{11}x_{1 0} \\&+ 220\,x_{0 1}^9x_{1 0}^3 + 792\,x_{0 1}^7x_{1 0}^5 + 792\,x_{0 1}^5x_{1 0}^7 + 220\,x_{0 1}^3x_{1 0}^9 + 12\,x_{0 1}x_{1 0}^{11} + x_{1 1}^{12}\,.
\end{aligned}
\end{align}
The corresponding partition function is
\begin{align}
\begin{aligned}
    Z_{B_1^{(12,2)}}
        =
        \frac{1}{|\eta|^{24}}\left[1 + 264\, q  + 576\, q^{\frac{1}{2}} \bar{q}^{\frac{1}{2}}+ 264\, \bar{q}  +\cdots\right]\,.
\end{aligned}
\end{align}
From the weight enumerator, it is clear that this code contains the all-ones vector and $\BF_4$-even.
Then, we can gauge the $\widehat{\BZ}_2$ symmetry, which is non-anomalous since $n=12\in4\BZ$.
The orbifold partition function is
\begin{align}
    \widehat{Z}^\CO_{B_1^{(12,2)}}
        =
        \frac{1}{|\eta|^{24}}\left[1 + 264\, q  + 264\, \bar{q}  +\cdots\right]\,.
\end{align}
The fermionized theory shows the following partition functions:
\begin{align}
\begin{aligned}
    \widehat Z_{B_1^{(12,2)}}^{\widetilde{\mathrm{R}}}
            &=
                576 \ , \\
        \widehat Z_{B_1^{(12,2)}}^\text{NS}
            &= \frac{1}{|\eta|^{24}}\left[1 + 264\, q  + 264 \,\bar{q}  + 
 2048 \,q^\frac{3}{2}  + 2048 \,\bar{q}^\frac{3}{2}  + 
 7944 \,q^2 + 69696\, q \bar{q}  + 
 7944 \bar{q}^2 +\cdots\right]\,,\\
    \widehat Z_{B_1^{(12,2)}}^\text{R}
            &= 576 + 98304 \,q + 98304 \,\bar{q}  + 
 2359296 \,q^2  + 16777216 \,q \bar{q}+ 
 2359296 \,\bar{q}^2 + \cdots \,.
\end{aligned}
\end{align}
The fermionic theory satisfies all the SUSY conditions and is expected to have supersymmetry.
Note that this fermionic theory is not based on the odd Leech lattice since the odd Leech lattice does not contain norm 1 vectors while the fermionic theory contains $528= 264+264$ norm 1 elements in Euclidean signature. Here we used the fact that the theta function $\sum q^{\frac{p_L^2}{2}}\bar{q}^{\frac{p_R^2}{2}}$ with Lorentzian signature reduces to the Euclidean one $\sum q^{\frac{p^2}{2}}$ by taking $q=\bar{q}$.

\subsection{$n=12,\, p=3$}
As a final example, we consider the B-form code associated with Pless symmetry code $\CP_{24}$.
The resulting code CFT has the non-anomalous $\BZ_2$ symmetry by which we can gauge to construct the orbifolded and fermionized theories.
Pless symmetry codes are shown to yield supersymmetric chiral CFTs in \cite{Gaiotto:2018ypj}, hence it is tempting to see if the B-form code CFT with the Pless symmetry code $\CP_{24}$ becomes a supersymmetric CFT after fermionization.

The bosonic partition function is
\begin{align}
    Z_{\CP_{24}} = \frac{1}{|\eta|^{24}}\left[1 + 4096\, q^\frac{3}{4} \bar{q}^\frac{3}{4} + 264\, \left(q^2 +\bar{q}^2\right)+ 
 16896\, \left(q^\frac{3}{2} \bar{q}^\frac{1}{2}+q^{\frac{1}{2}} \bar{q}^\frac{3}{2}\right) + 63936\, q \bar{q}  + \cdots\right]\,.
\end{align}
Note that if one takes $q=\bar{q}$, then the lattice theta function becomes that of the odd Leech lattice in Euclidean signature, which means that the bosonic theory is based on the odd Leech lattice equipped with the Lorentzian metric.
On the other hand, the orbifold partition function is
\begin{align}
    Z^\CO_{\CP_{24}} = \frac{1}{|\eta|^{24}}\left[1 + 24\, \left(q +\bar{q}\right)   + 264\, \left(q^2 +\bar{q}^2\right)+ 
 33792\, \left(q^\frac{3}{2}\bar{q}^\frac{1}{2}+q^{\frac{1}{2}}\bar{q}^\frac{3}{2}\right)   + 127296\, q\bar{q} + 
  \cdots\right]
\end{align}
The fermionized partition functions are
\begin{align}
    \begin{aligned}Z^{\widetilde{\mathrm{R}}}_{\CP_{24}}
        &=
            -24\,\left(q^{\frac{1}{2}}\Bar{q}^{-\frac{1}{2}}+\,q^{-\frac{1}{2}}\Bar{q}^{\frac{1}{2}}\right) + 4096\,q^{\frac{1}{4}}\bar{q}^{\frac{1}{4}}
            -288\,\left(q^{\frac{3}{2}}\Bar{q}^{-\frac{1}{2}}+\,q^{-\frac{1}{2}}\Bar{q}^{\frac{3}{2}}\right)-16896\,\left(q+\Bar{q}\right)+\cdots\,,\\
    Z^{\text{R}}_{\CP_{24}}
        &=
            24\,\left( q^{\frac{1}{2}}\Bar{q}^{-\frac{1}{2}} + q^{-\frac{1}{2}}\Bar{q}^{\frac{1}{2}}\right) + 288\,\left( q^{\frac{3}{2}}\Bar{q}^{-\frac{1}{2}}+q^{-\frac{1}{2}}\Bar{q}^{\frac{3}{2}}\right) + 16896\,\left( q+\Bar{q}\right) +\cdots\,,\\
    Z^{\text{NS}}_{\CP_{24}}
        &=\frac{1}{|\eta|^{24}}\left[
            1 + 264\, \left(q^2  + \Bar{q}^2\right) + 16896\, \left(q^\frac{3}{2} \Bar{q}^\frac{1}{2} + q^\frac{1}{2} \Bar{q}^\frac{3}{2}\right) + 49152\, \left(q^\frac{5}{4} \Bar{q}^\frac{3}{4} + q^\frac{3}{4} q^\frac{5}{4} \right) + 63936\, q \Bar{q}  +\cdots\right]\,.
            \end{aligned}
\end{align}
These partition functions show that this theory does not admit supersymmetry.

\section{Discussion}\label{ss:discussion}

This paper provides the construction of Narain CFTs from quantum subsystem codes.
Subsystem codes are a generalization of stabilizer codes by adding a new ingredient called gauge qudits.
While a stabilizer code is characterized by an abelian stabilizer group $\CS$, a subsystem code is by a non-abelian gauge group $\CG$.
We utilize a map from a gauge group $\CG$ of a subsystem $[[n,k,k]]_p$ code to a classical $[2n,n]_p$ code and Construction A of Lorentzian lattices from classical codes.
By regarding the Lorentzian lattice as a momentum lattice, we obtain a Narain CFT defined by a set of vertex operators.

Having the classification of code CFTs in mind, we enumerate the B-form subsystem codes characterized by weighted oriented graphs and investigate the corresponding code CFTs as well as their orbifolded and fermionized theories by gauging the $\BZ_2$ symmetries.
Within the B-form code CFTs, we find various bosonic CFTs that are self-dual under the orbifold and fermionic CFTs with supersymmetry.
Incidentally, the classical codes corresponding to the B-form subsystem codes include an interesting family of (an infinite number of) codes such as Pless symmetry codes and quadratic double circulant codes.
While we are not aware of their peculiarities in this paper, they may have potential applications in further investigations of Narain CFTs in future.

In section~\ref{ss:fermion}, we explicitly construct $\CN=1,2$ supercurrents in specific cases.
In this construction, it is essential for the theory not to contain vertex operators with conformal dimension $(h,\bar{h}) = (1,0),(2,0)$. Otherwise, we need to take care of relative phases arising from OPEs between vertex operators, which are characterized by cocycle factors (see for example~\cite{Polchinski:1998rq}).
Recently, it has been pointed out that this subtlety can be resolved by using quantum error-correcting condition~\cite{Harvey:2020jvu}, and the authors of~\cite{Moore:2023zmv} applied the method to the manifest construction of a supercurrent in ``Beauty and the Beast" $\CN=1$ SCFT~\cite{Dixon:1988qd}.
Making use of it, we may extend our analysis in section~\ref{ss:fermion} to a broader class of fermionic CFTs satisfying the SUSY conditions and prove the existence of supersymmetry by identifying the supercurrents.

In section~\ref{ss:enumeration}, we find that some bosonic CFTs are self-dual under gauging a global $\BZ_2$ symmetry.
In a recent discussion about generalized symmetry, this is evidence for the theory to admit non-invertible symmetry, whose fusion rule is beyond the group-theoretical framework.
There are several constructions of non-invertible symmetries (see, for example~\cite{Shao:2023gho}). 
Among them, the half gauging is a powerful technique for constructing non-invertible duality defects.
The half gauging construction separates the spacetime into two parts: the original theory lives in the left region, while the gauge theory lives in the right region.
To construct duality defects, we need a precise duality connecting the left theory to the right one.
In our cases, the duality would be given by T-duality as seen in $c=2$ Narain CFTs in \cite{Nagoya:2023zky}.
It would be interesting to construct duality defects associated with the $\BZ_2$ symmetries we have considered.

In~\cite{Dymarsky:2020qom,Kawabata:2022jxt}, Narain CFTs are constructed from stabilizer codes.
In the construction, stabilizer codes that are able to give modular-invariant theories do not have logical qudits.
Such a class of stabilizer codes has a one-dimensional code subspace and cannot encode any information.
In contrast, our construction utilizes subsystem $[[n,k,k]]_p$ codes and they have $k$ logical and $k$ gauge qudits.
Correspondingly, those subsystem codes can store the information of $k$ qudits in $n$ qudits.
We leave it as a future work to explore the roles and implications of logical qudits in the context of code CFTs.

\acknowledgments
The work of T.\,N. was supported in part by the JSPS Grant-in-Aid for Scientific Research (C) No.19K03863, Grant-in-Aid for Scientific Research (A) No.\,21H04469, and
Grant-in-Aid for Transformative Research Areas (A) ``Extreme Universe''
No.\,21H05182 and No.\,21H05190, and Grant-in-Aid for Scientific Research (B) No.\,24K00629.
The work of K.\,K. was supported by FoPM, WINGS Program, the University of Tokyo, and by JSPS KAKENHI Grant-in-Aid for JSPS fellows Grant No.\,23KJ0436.

\appendix

\section{A family of B-form codes}
\label{ap:Pless_QDC}

\subsection{Pless symmetry codes}

For any prime $q =5~(\text{mod}~ 6)$, Pless symmetry code $\CP_{2q+2}$ \cite{pless1972symmetry,macwilliams1977theory,conway2013sphere} is defined as the (Euclidean self-dual) $[2q+2, q+1]$ code over $\BF_3$ with generator matrix $G=[\, I\, |\, B\,]$ where $S$ is the $(q+1)\times (q+1)$ matrix (with rows and columns labeled $\infty, 0, 1, \cdots, q-1$) given by
\NiceMatrixOptions
{
    custom-line ={command= H, tikz= dashed, width= 1mm}, 
    custom-line = {letter= I, tikz= dashed, width= 1mm}, 
}
\begin{align}
    B 
        =
        \begin{bNiceArray}{cI cccc}[first-row, last-col, cell-space-top-limit=3pt, extra-margin=4pt]
           \infty   & 0     & 1     & \cdots & q-1  &   \\
            0       & ~1~     & ~1~     & \cdots & ~1~    &  \infty \\
            \H
           \epsilon &       &       &        &      &  0  \\
           \epsilon &       &       &        &      &  1  \\
            \vdots  &       &       &   C    &      &  \vdots \\
           \epsilon &       &       &        &      &  q-1  \\
        \end{bNiceArray} \ , \qquad 
    \epsilon
        =
        \begin{cases}
            +1 \qquad \text{for}\quad q = 4k + 1 \\
            -1 \qquad \text{for}\quad q = 4k +3 
        \end{cases} \ .
\end{align}
$C = (c_{ij})~(i,j=0,1,\cdots, q-1)$ is a circulant matrix with
\begin{align}
    c_{ij}
        =
        \begin{cases}
            0 \quad \text{if}~ i = j \\
            1 \quad \text{if}~j - i~ \text{is a square mod}~ q ~ (i\neq j) \\
            -1 \quad \text{if}~j - i~ \text{is not a square mod}~ q ~ (i\neq j)
        \end{cases}
\end{align}
Pless symmetry codes satisfy the followings:
\begin{itemize}
    \item $B\,B^T = - I~(\text{mod}~3)$
    \item $B = B^T = - B^{-1}$ if $q=4k+1$ and $B = -B^T = B^{-1}$ if $q=4k+3$
    \item All weights in $\CP_{2q+2}$ are divisible by $3$
    \item $\CP_{2q+2}$ contains the all-ones vector \cite{conway2013sphere}
\end{itemize}
Thus, when $q = 12\,k + 11~(k=0,1,\cdots)$ is prime, Pless symmetry codes satisfy the condition \eqref{B-form_CFT_condition} and there exist code CFTs associated with them.
Hence, Pless symmetry codes correspond to $[[n,\frac{n}{2},\frac{n}{2}]]_3$ subsystem codes with $n=12\,\BZ_+$.

Let us consider the case with $q=11$, which gives the $[24, 12, 9]_3$ code $\CP_{24}$ with the generator matrix \cite{leon1981ternary}:
\begin{align}
    G
        =
        \left[
        \begin{array}{cccccccccccc|rrrrrrrrrrrr}
            ~1 & & & & & & & & & & & &  0 & 1 & 1 &  1 &  1 &  1 &  1 &  1 &  1 &  1 &  1 &  1 \\
              & ~1 & & & & & & & & & & & -1 & 0 & 1 & -1 & 1 & 1 & 1 & -1 & -1 & -1 & 1 & -1 \\
            & & ~1 & & & & & & & & & & ~-1 & -1 & 0 & 1 & -1 & 1 & 1 & 1 & -1 & -1 & -1 & 1 \\
            & & & ~1 & & & & & & & & & -1 &  1 & -1 & 0 & 1 & -1 & 1 & 1 & 1 & -1 & -1 & -1 \\
            & & & & ~1 & & & & & & & &  -1 & -1 & 1 & -1 & 0 & 1 & -1 & 1 & 1 & 1 & -1 & -1 \\
            & & & & & ~1 & & & & & & & -1 & -1 & -1 & 1 & -1 & 0 & 1 & -1 & 1 & 1 & 1 & -1 \\
            & & & & & & ~1 & & & & & & -1 & -1 & -1 & -1 & 1 & -1 & 0 & 1 & -1 & 1 & 1 & 1 \\ 
            & & & & & & & ~1 & & & & & -1 & 1 & -1 & -1 & -1 & 1 & -1 & 0 & 1 & -1 & 1 & 1 \\
            &  & & & & & & & ~1 & & & & -1 & 1 & 1 & -1 & -1 & -1 & 1 & -1 & 0 & 1 & -1 & 1\\
            & &  & & & & & & & ~1 &  & & -1 & 1 & 1 & 1 & -1 & -1 & -1 & 1 & -1 & 0 & 1 & -1 \\
            & & &  & & & & & & & ~1 & & -1 & -1 & 1 & 1 & 1 & -1 & -1 & -1 & 1 & -1  & 0 & 1\\
            & & & &  & & & & &  &  & ~1~ & -1  &1 & -1 & 1 & 1 & 1 & -1 & -1 & -1 & 1 & -1 & 0 \\
        \end{array}
        \right]
\end{align}
The Construction A lattice $\Lambda(\CP_{24})$ for the Pless symmetry code $\CP_{24}$ is an odd self-dual lattice of 24 dimensions, whose shortest vector has square-length $3$ \cite{Gaiotto:2018ypj}.
Thus, $\Lambda(\CP_{24})$ is the odd Leech lattice $O_{24}$.

\subsection{Quadratic double circulant codes}
\label{ap:QDC_code}

The generalization of the Pless symmetry codes to a finite field $\BF_l$ is given and called a quadratic double circulant code in \cite{gaborit2002quadratic}.
For prime $q$, there are two classes of quadratic double circulant codes whose generator matrices are given by
\begin{align}
    \CP_{2q}(r, s, t) 
            &=
            \left[
            \begin{array}{c|c}
        	~I_q ~& ~Q_q(r,s,t)~ 
            \end{array}
        \right] \ , \\
    \CP_{2q+2}(r, s, t) 
            &=
            \begin{bNiceArray}{cI cccc | cI cccc}[first-row, last-col, cell-space-top-limit=3pt, extra-margin=4pt]
           & & & & & \infty   & 0     & 1     & \cdots & q-1  &   \\
            1 & ~0~ & ~0~ & \cdots & ~0~ & ~r~       & ~1~     & ~1~     & \cdots & ~1~    &  \infty \\
            \H
           0 & & & & & \epsilon &       &       &        &      &  0  \\
           0 & & & & & \epsilon &       &       &        &      &  1  \\
            \vdots & & & I_q & &\vdots  &       &       &   Q_q(r,s,t)    &      &  \vdots \\
           0 & & & & & \epsilon &       &       &        &      &  q-1  \\
        \end{bNiceArray} \\
        &\equiv 
            \left[
            \begin{array}{c|c}
        	~I_{q+1} ~& ~S_{q+1}(r,s,t)~ 
            \end{array}
        \right]
\end{align}
where $Q_q(r,s,t)$ is a circulant matrix with the entries, called the Paley (or Jacobsthal) matrix, defined by
\begin{align}
    \left(Q_q(r,s,t)\right)_{ij}
        =
        \begin{cases}
            r & \text{if}~ i=j \\
            s & \text{if $j-i$ is a square mod $q$ ($i\neq j$)} \\
            t & \text{if $j-i$ is a not square mod $q$ ($i\neq j$)}
        \end{cases}
\end{align}
and
\begin{align}
    \epsilon
        &=
        \begin{cases}
            +1 \qquad \text{for}\quad q = 4k + 1 \\
            -1 \qquad \text{for}\quad q = 4k + 3 
        \end{cases} \ .
\end{align}

In what follows, we set the parameters $r,s,t$ to special values for simplicity:
\begin{align}
    r = 0 \ , \qquad s = 1 \ , \qquad t = -1 \ ,
\end{align}
and denote the matrices as $Q_q = Q_q(0,1,-1), S_{q+1} = S_{q+1}(0,1,-1)$ and similarly the $[2q,q]_l$ and $[2q+2, q+1]_l$ codes as $\CP_{2q}$ and $\CP_{2q+2}$, respectively.

When $q$ is a power of an odd prime, one can show the following:

\begin{itemize}
    \item 
        $
            Q_q^2 = q\,I_q - J_q
        $
        where $J_q$ is the $q\times q$ all one matrix.
    \item $Q_q\,J_q = J_q\,Q_q = 0$
    \item $S_{q+1}\, S_{q+1}^T = q\,I_{q+1}$

    \item $Q_q = Q_q^T, ~ S_{q+1} = S_{q+1}^T$ if $q=4k+1$ and $Q_q = -Q_q^T, ~ S_{q+1} = -S_{q+1}^T$ if $q=4k+3$.

\end{itemize}

\bibliographystyle{JHEP}
\bibliography{QEC_CFT_prime_power}
\end{document}